\documentclass[aps,onecolumn,letterpaper,oneside,preprint,tightenlines,draft,%
nobibnotes,nofootinbib,amsfonts,amssymb,amsmath,eqsecnum,showpacs,%
showkeys]{revtex4}

\newcommand{\cN}{\mathcal{N}}
\newcommand{\cV}{\mathcal{V}}
\newcommand{\cW}{\mathcal{W}}
\newcommand{\bA}{\boldsymbol{A}}
\newcommand{\bB}{\boldsymbol{B}}
\newcommand{\bC}{\boldsymbol{C}}
\newcommand{\bH}{\boldsymbol{H}}
\newcommand{\bI}{\boldsymbol{I}}
\newcommand{\bJ}{\boldsymbol{J}}
\newcommand{\bP}{\boldsymbol{P}}
\newcommand{\bQ}{\boldsymbol{Q}}
\newcommand{\bT}{\boldsymbol{T}}
\newcommand{\bU}{\boldsymbol{U}}
\newcommand{\bV}{\boldsymbol{V}}
\newcommand{\bcV}{\boldsymbol{\cV}}
\newcommand{\be}{\boldsymbol{e}}
\newcommand{\bw}{\boldsymbol{w}}
\newcommand{\tH}{\tilde{H}}
\newcommand{\tbH}{\tilde{\bH}}
\newcommand{\tP}{\tilde{P}}
\newcommand{\tbP}{\tilde{\bP}}
\newcommand{\tcV}{\tilde{\cV}}
\newcommand{\tbcV}{\tilde{\bcV}}
\newcommand{\tbw}{\tilde{\bw}}
\newcommand{\tbpsi}{\tilde{\boldsymbol{\psi}}}
\newcommand{\bbN}{\mathbb{N}}

\newcommand{\rme}{\mathrm{e}}
\newcommand{\rmd}{\mathrm{d}}
\newcommand{\del}{\partial}
\newcommand{\Llra}{\Longleftrightarrow}
\newcommand{\lra}{\leftrightarrow}
\newcommand{\rmA}{\mathrm{A}}
\newcommand{\rmB}{\mathrm{B}}
\newcommand{\rmT}{\mathrm{T}}

\newcommand{\braket}[1]{\bigl\langle{#1}\bigr\rangle}
\newcommand{\fsl}{\mathfrak{sl}}

\newcommand{\fq}{\mathfrak{q}}
\newcommand{\fosp}{\mathfrak{osp}}

\begin{document}


%
%

\title{Construction of Quasi-solvable Quantum Mechanical Matrix Models:
 Lie Superalgebra v.s.\ $\boldsymbol{\cN}$-fold Supersymmetry}
\author{Toshiaki Tanaka}
\email{tanaka.toshiaki@ocha.ac.jp}
\affiliation{Department of Physics,
 Faculty of Science, Ochanomizu University,
 2-1-1 Ohtsuka, Bunkyo-ku, Tokyo 112-8610, Japan}


\begin{abstract}

We construct quasi-solvable quantum mechanical matrix models by employing two different methods, the one is universal enveloping algebra of Lie superalgebra and the other is $\cN$-fold supersymmetry. For the former we examine the $\fq(2)$ and $\fosp(2/2)$ Lie-superalgebraic quasi-solvable matrix operators in the literature, and then compare them with the corresponding $\cN$-fold supersymmetric matrix systems. In the $\fq(2)$ case, Lie-superalgebraic construction and the intertwining relation lead to the identical result. In the $\fosp(2/2)$ case, however, some novel features emerge due to the difference in dimension of linear spaces which consist of the two-component invariant subspace. In both cases, the closure of $\cN$-fold superalgebra imposes stronger constraint on the admissible form of the systems and the concept of conjugation plays a key role in the formulation.

\end{abstract}






\maketitle

\section{Introduction}
\label{sec:intro}

Since the discovery of quasi-exactly solvable models~\cite{TU87} and its underlying $\fsl(2)$ Lie-algebraic structure~\cite{Tu88}, many applications of similar Lie-algebraic methods were worked out in 1990s~\cite{Us94}. Then, the equivalence between weak quasi-solvability and $\cN$-fold supersymmetry (SUSY) proved in~\cite{AST01b} has further enhanced systematic investigations into quasi-solvable quantum mechanical systems~\cite{Ta09}. In comparison to ordinary scalar Schr\"{o}dinger operator, however, the progress in studying analytic solutions to matrix differential operators does not seem so significant although such systems appear in various physical contexts such as Dirac, Pauli, and multi-channel Schr\"{o}dinger equations. Taking Dirac equation as an example, we would find that most of the approaches in the literature has relied on the procedure to build up potential terms in Dirac equation such that it is reducible to one of the known (quasi-)exactly solvable Schr\"{o}dinger equations.

The fact is that there have been several works on constructing quasi-solvable matrix systems by extending Lie algebra, which have been widely utilized in the scalar case, to Lie superalgebra and beyond. The first such attempt appeared in~\cite{ST89,Sh89} where a differential realization of the Lie superalgebra $\fosp(2/2)$ was used to construct $2\times 2$ matrix models. For more details of this approach, see also~\cite{Tu92b,Tu94}. By generalizing the two-component $\fosp(2/2)$ module, more complicated algebra beyond Lie superalgebra was considered in~\cite{BK94,FGR97}. The latter formulation was further extended to $2\times 2$ matrix models of multi variables in~\cite{BN98}. There were Lie algebraic approaches, construction of $n\times n$ matrix quasi-solvable differential operators was formulated by using matrix representations of the Lie algebra $\fsl(2)$ in~\cite{Zh97a}, and similar idea was followed by considering $\mathfrak{o}(2,2)$, which is extendable to Lie superalgebra, for $n=2$ in~\cite{SZ99}. Another approach to construct $2\times 2$ matrix quasi-solvable models was made by employing the Lie superalgebra $\fosp(2/1)$ in~\cite{SK97} and by $\fq(2)$ in~\cite{DJ01}. An application of the latter was appeared in~\cite{BH02}. For other approaches, see e.g.~\cite{Bri00,BH01}.

In our previous paper~\cite{Ta12a}, on the other hand, we successfully established the framework of $\cN$-fold SUSY in quantum mechanical matrix models in such a way that their weak quasi-solvability is automatically guaranteed. We also constructed explicitly the most general 2-fold SUSY $2\times 2$ Hermitian matrix systems. Those achievements suggests us the possibility that the existing Lie-superalgebraic quasi-solvable models mentioned above would be reproduced, fully or partially, by the $\cN$-fold SUSY formulation. In addition, it is quite interesting to see whether we can discover unexpected novel aspects which have not been observed in the conventional Lie-superalgebraic approach. In this respect, it should be reminded that the equivalence between weak quasi-solvability and $\cN$-fold SUSY has not been confirmed nor disproved in the matrix case, which was proved to be true only in the scalar case. Indeed, the analysis in the previous paper~\cite{Ta12a} indicated the breakdown of the equivalence for matrix systems.

In this paper, we construct quasi-solvable quantum mechanical matrix models by employing the two methods, the one is based on universal enveloping algebra of Lie superalgebra and the other is $\cN$-fold SUSY. For the purpose, we slightly sophisticate the formulation of $\cN$-fold SUSY matrix systems in~\cite{Ta12a} by introducing the concept of conjugation which turns out to play an important role in clarifying the difference between the intertwining relation and the algebraic constraint. The Lie-superalgebraic models we shall treat in this paper are the $\fq(2)$ one realized by the second representation in~\cite{DJ01} and the $\fosp(2/2)$ model in~\cite{Tu92b}. We then make attempts to construct the corresponding $\cN$-fold SUSY systems by applying the developed formulation. The application consists of two processes, namely, to solve the intertwining relation and to solve the algebraic constraint. We compare the results in each step with the Lie-superalgebraic models and examine their relation.

We organize the paper as follows. In Section~\ref{sec:Nf0}, we develop the formulation of $\cN$-fold SUSY in quantum mechanical systems in a more general way than the one in~\cite{Ta12a} by introducing the notion of conjugation. We then discuss in detail consequences of requiring the invariance of a system under the conjugation. In Section~\ref{sec:q2}, we review the $\fq(2)$ Lie-superalgebraic quasi-solvable matrix model in~\cite{DJ01} and apply it to our case. The most general quasi-solvable matrix model which meets the form in our consideration and can be constructed by the universal enveloping algebra of $\fq(2)$ is presented and examined. In Section~\ref{sec:Nf1}, we apply the framework of $\cN$-fold SUSY in Section~\ref{sec:Nf0} to construct a system which would have direct relation with the $\fq(2)$ model in Section~\ref{sec:q2}. We show that the system just obtained by solving the intertwining relation contains a component which is exactly identical to the $\fq(2)$ model. We further show that the requirement of the transposition symmetry and of the closure of $\cN$-fold superalgebra lead to almost the same consequence and that the resultant model is equivalent to a scalar type A $\cN$-fold SUSY system. In Section~\ref{sec:osp22}, we review the $\fosp(2/2)$ Lie-superalgebraic construction of quasi-solvable matrix models in~\cite{Tu92b} and apply it to our case. We present the most general quasi-solvable model constructed from the universal enveloping algebra of $\fosp(2/2)$ which falls into our considering form. In Section~\ref{sec:Nf2}, we employ the $\cN$-fold SUSY formulation for the attempt to reconstruct the $\fosp(2/2)$ model in Section~\ref{sec:osp22}. We discuss in detail novel features which do not emerge in the $\fq(2)$ case and their underlying causes. As a consequence, the $\cN$-fold supercharge admits two additional functions in its components and its kernel extends the originally considered invariant subspace. We show as a special case that the system turns to be composed of one scalar type B and one scalar type A $\cN$-fold SUSY systems when one of the extra functions is trivial. For the general case, we cannot solve the intertwining relations for an arbitrary $\cN$ and investigate only the $\cN=1$ and $2$ cases. In both the two cases, we can obtain by solving the intertwining relation particular cases of the $\fosp(2/2)$ model whose forms are restricted by the extra invariant vector. In Section~\ref{sec:discus}, we summarize the results and provide various prospects and future issues to be pursued.

\section{$\cN$-fold SUSY QM Matrix Models}
\label{sec:Nf0}

$\cN$-fold SUSY for quantum mechanical matrix operators was successfully formulated in Ref.~\cite{Ta12a}. Here we shall slightly generalize and sophisticate its formulation. Suppose we have a kind of conjugate operation for matrix differential operators (e.g., Hermitian conjugate, transposition), denoted by $\#$, satisfying $(\bA^{\#})^{\#}=\bA$ and $(\bA\bB)^{\#}=\bB^{\#}\!\bA^{\#}$ for any $\bA$ and $\bB$, and that a pair of matrix differential operators $\bH^{\pm}$ is invariant under it, $(\bH^{\pm})^{\#}=\bH^{\pm}$. Let $\bP_{\cN}^{-}$ be a matrix linear differential operator of order $\cN$. Then, we shall say the system $(\bH^{\pm}, \bP_{\cN}^{-})$ is $\cN$-fold SUSY if the pair $\bH^{\pm}$ is intertwined by $\bP_{\cN}^{-}$
\begin{align}
\bP_{\cN}^{-}\bH^{-}-\bH^{+}\bP_{\cN}^{-}=0,
\label{eq:Nf1}
\end{align}
and in addition satisfies
\begin{align}
\bP_{\cN}^{\mp}\bP_{\cN}^{\pm}=2^{\cN}\left[ (\bH^{\pm}+\bC_{0})^{\cN}+\sum_{k=1}^{\cN-1}\bC_{k}(\bH^{\pm}+\bC_{0})^{\cN-k-1}\right],
\label{eq:Nf2}
\end{align}
where $\bP_{\cN}^{+}=(\bP_{\cN}^{-})^{\#}$, and $\bC_{k}$ ($k=0,\dots,\cN-1$) are constant matrices satisfying
\begin{align}
\bC_{0}^{\#}=\bC_{0},\quad (\bH^{\pm}+\bC_{0})\bC_{k}^{\#}=\bC_{k}(\bH^{\pm}+\bC_{0})\quad (k=1,\dots,\cN-1).
\label{eq:Ccon}
\end{align}
They guarantee the invariance of $\bP_{\cN}^{\mp}\bP_{\cN}^{\pm}$ under the conjugation $\#$ in (\ref{eq:Nf2}). We note that (\ref{eq:Nf1}) automatically ensures its conjugate relation:
\begin{align}
\bP_{\cN}^{+}\bH^{+}-\bH^{-}\bP_{\cN}^{+}=-\left( \bP_{\cN}^{-}\bH^{-}-\bH^{+}\bP_{\cN}^{-}\right)^{\#}=0.
\label{eq:Nf1'}
\end{align}
Weak quasi-solvability of $\bH^{\pm}$ automatically follows from (\ref{eq:Nf1}) and (\ref{eq:Nf1'}):
\begin{align}
\bH^{\pm}\ker\bP_{\cN}^{\pm}\subset\ker\bP_{\cN}^{\pm}.
\end{align}
Hence, $\cN$-fold SUSY provides a sufficient condition for weak quasi-solvablity. The converse has been proved only for the scalar case \cite{AST01b}, and it may be false in the matrix case. Thus, the second requirement (\ref{eq:Nf2}) may cause a stricter restriction on a possible class of weakly quasi-solvable matrix systems.

Until now, the form of matrix operators $\bH^{\pm}$ is arbitrary, but hereafter we shall restrict them to $n\times n$ matrix Schr\"{o}dinger operators possessing the following form:
\begin{align}
\bH^{\pm}=-\frac{1}{2}\bI_{n}\frac{\rmd^{2}}{\rmd q^{2}} +\bU^{\pm}(q)\frac{\rmd}{\rmd q}+\bV^{\pm}(q).
\label{eq:Hpm1}
\end{align}
A natural choice of the conjugation $\#$ is Hermitian conjugation, usually denoted by superscript $\dagger$. For a matrix linear differential operator under consideration, it is taken both in the functional space on which differential operators act and in the $n$-dimensional linear space on which matrices act. Then, the self-conjugate simply means Hermiticity of the matrix Hamiltonians $(\bH^{\pm})^{\dagger}=\bH^{\pm}$, and in the case of (\ref{eq:Hpm1}) it leads to the following set of conditions:
\begin{align}
(\bU^{\pm})^{\dagger}(q)=-\bU^{\pm}(q),\qquad (\bV^{\pm})^{\dagger}(q)=\bV^{\pm}(q)-\bU^{\pm\prime}(q).
\label{eq:Herm}
\end{align}
The Hermitian construction was actually employed in Ref.~\cite{Ta12a}.

Another possible choice of the conjugation $\#$ is transposition, both in the functional space on which differential operators act and in the $n$-dimensional linear space on which matrices act. We shall employ superscript $\rmT$ to denote the transpositions commonly in the whole or in either spaces. In this choice of the conjugation $\#=\rmT$, the matrices $\bU^{\pm}(q)$ and $\bV^{\pm}(q)$ in (\ref{eq:Hpm1}) must satisfy
\begin{align}
(\bU^{\pm})^{\rmT}(q)=-\bU^{\pm}(q),\qquad (\bV^{\pm})^{\rmT}(q)=\bV^{\pm}(q)-\bU^{\pm\prime}(q),
\label{eq:tps1}
\end{align}
to guarantee the invariance of $\bH^{\pm}$ under the transposition. In particular, $\bU^{\pm}(q)$ must be anti-symmetric and its diagonal elements are zero.

There are still other candidates for the conjugation $\#$, for example, as the combination of the transposition in the functional space and inversion in the $n$-dimensional linear space, provided that each matrix coefficient of linear differential operators involved in a whole system is regular and has its own inverse. In this case, the invariance of $\bH^{\pm}$ in (\ref{eq:Hpm1}) under the conjugation $\#$ reads instead as
\begin{align}
(\bU^{\pm})^{-1}(q)=-\bU^{\pm}(q),\qquad (\bV^{\pm})^{-1}(q)=\bV^{\pm}(q)-\bU^{\pm\prime}(q).
\label{eq:inver}
\end{align}
In this work, we shall hereafter restrict our consideration to the transposition as the conjugation, $\#=\rmT$, since it seems to be simpler and generalization to other cases would be straightforward.

The most general form of a matrix linear differential operator $\bP_{\cN}^{-}$ to be considered is
\begin{align}
\bP_{\cN}^{-}=\bI_{n}\frac{\rmd^{\cN}}{\rmd q^{\cN}}+\sum_{k=0}^{\cN-1}\bw_{k}(q)\frac{\rmd^{k}}{\rmd q^{k}},
\label{eq:Pm1}
\end{align}
and its conjugation thus reads as
\begin{align}
\bP_{\cN}^{+}=(\bP_{\cN}^{-})^{\rmT}=(-1)^{\cN}\bI_{n}\frac{\rmd^{\cN}}{\rmd q^{\cN}}+\sum_{k=0}^{\cN-1}(-1)^{k}\frac{\rmd^{k}}{\rmd q^{k}}\bw_{k}^{\rmT}(q).
\end{align}
To construct an $\cN$-fold SUSY system in practice, it is often more convenient to work in a `gauged' $z$-space rather than in the physical $q$-space. The transition from the $q$-space to a $z$-space is accomplished by a change of variable $z=z(q)$ and a gauge transformation. For matrix operators the gauge transformation can be matrix-valued, but in this paper we shall restrict ourselves to a scalar gauge transformation. It means that for a matrix operator $\bA$ in the $q$-space we would have the corresponding gauged one $\tilde{\bA}$ connected by a scalar potential $\cW_{\cN}^{-}$ as $\tilde{\bA}=\rme^{\cW_{\cN}^{-}}\bA\,\rme^{-\cW_{\cN}^{-}}$. Hence, the gauged matrix Hamiltonians $\tbH^{\pm}$ would result in
\begin{align}
\tbH^{\pm}=\rme^{\cW_{\cN}^{-}}\bH^{\pm}\,\rme^{-\cW_{\cN}^{+}}=-\bA(z)\frac{\rmd^{2}}{\rmd z^{2}}-\bB^{\pm}(z)\frac{\rmd}{\rmd z}-\bC^{\pm}(z).
\label{eq:tHpm1}
\end{align}
Substituting (\ref{eq:Hpm1}) into (\ref{eq:tHpm1}), we get the relations among the quantities in the $q$-space and in the $z$-space:
\begin{align}
z'(q)^{2}\bI_{n}&=2\bA(z(q))=2A(z(q))\bI_{n},
\label{eq:rel1}\\
z'(q)\bU^{\pm}(q)&=\left( \frac{A'(z)}{2}-2\frac{\rmd\cW_{\cN}^{-}}{\rmd z}A(z)\right) \bI_{n}-\bB^{\pm}(z)\biggr|_{z=z(q)},
\label{eq:rel2}\\
\bV^{\pm}(q)&=-\left[ \frac{\rmd^{2}\cW_{\cN}^{-}}{\rmd z^{2}}+\left(\frac{\rmd\cW_{\cN}^{-}}{\rmd z}\right)^{2}\right] A(z)\bI_{n}-\frac{\rmd\cW_{\cN}^{-}}{\rmd z}\bB^{\pm}(z)-\bC^{\pm}(z)\biggr|_{z=z(q)}.
\label{eq:rel3}
\end{align}
In particular, the first relation (\ref{eq:rel1}) determines the change of variable $z=z(q)$. Furthermore, it forces $\bA(z)$ proportional to $\bI_{n}$.
Under those conditions, the gauged $\cN$-fold supercharge components read as
\begin{align}
\tbP_{\cN}^{-}&=z'(q)^{\cN}\left[ \bI_{n}\frac{\rmd^{\cN}}{\rmd z^{\cN}}+\sum_{k=0}^{\cN-1}\tbw_{k}(z)\frac{\rmd^{k}}{\rmd z^{k}}\right],\\
\tbP_{\cN}^{+}&=z'(q)^{\cN}\left[ \bI_{n}\left(-\frac{D}{D z}\right)^{\cN}+\sum_{k=0}^{\cN-1}\left(-\frac{D}{D z}\right)^{k}\tbw_{k}^{\rmT}(z)\right],
\end{align}
where the `covariant' derivative is introduced as
\begin{align}
\frac{D}{D z}=\frac{\rmd}{\rmd z}+\frac{(\cN-1)A'(z)}{2A(z)}-2\frac{\rmd\cW_{\cN}^{-}}{\rmd z}.
\label{eq:covd1}
\end{align}
Next, we shall examine the effect of the transposition symmetry (\ref{eq:tps1}) in the $q$-space on the quantities in the $z$-space. From the second relation (\ref{eq:rel2}), the transformation rule of $\bB^{\pm}$ under the transposition reads as
\begin{align}
(\bB^{\pm})^{\rmT}(z)=-\bB^{\pm}(z)+\left( A'(z)-4\frac{\rmd\cW_{\cN}^{-}}{\rmd z}A(z)\right) \bI_{n}.
\label{eq:tB1}
\end{align}
It suggests that the matrix $\bB^{\pm}$ can be decomposed into a diagonal part which is a multiple of $\bI_{n}$ and an anti-symmetric part as
\begin{align}
\bB^{\pm}(z)=B(z)\bI_{n}+\bB_{A}^{\pm}(z),
\label{eq:tB2}
\end{align}
where
\begin{align}
2B(z)=A'(z)-4\frac{\rmd\cW_{\cN}^{-}}{\rmd z}A(z),\qquad 
\left(\bB_{A}^{\pm}\right)^{\rmT}(z)=-\bB_{A}^{\pm}(z).
\label{eq:tB3}
\end{align}
Expressed in components, it is equivalent to
\begin{align}
2B_{11}^{\pm}=\dots=2B_{nn}^{\pm}=2B=A'-4\frac{\rmd\cW_{\cN}^{-}}{\rmd z}A,\quad 
B_{ji}^{\pm}=-B_{ij}^{\pm}=-B_{Aij}^{\pm}\quad (i\neq j),
\label{eq:tB4}
\end{align}
where $B_{ij}^{\pm}=(\bB^{\pm})_{ij}$ and $B_{Aij}^{\pm}=(\bB_{A}^{\pm})_{ij}$ are the $(i,j)$th component of $\bB^{\pm}$ and $\bB_{A}^{\pm}$, respectively.
This means that the matrix $\bB^{\pm}$ must have the form indicated in (\ref{eq:tB4}) in order that the operator (\ref{eq:tHpm1}) could transform back to a matrix Schr\"{o}dinger operator (\ref{eq:Hpm1}) with transposition symmetry. The first formula in (\ref{eq:tB4}) determines the gauge potential $\cW_{\cN}^{-}$ for given $A(z)$ and $B(z)$:
\begin{align}
\frac{\rmd\cW_{\cN}^{-}}{\rmd z}=\frac{A'(z)-2B(z)}{4A(z)}.
\label{eq:WN-1}
\end{align}
Next, taking the transposition of (\ref{eq:rel3}) and then applying the second formula in (\ref{eq:tps1}), (\ref{eq:tB1}), and (\ref{eq:WN-1}), we have
\begin{align}
\bU^{\pm\prime}(q)=(\bC^{\pm})^{\rmT}(z)-\bC^{\pm}(z)-\frac{A'(z)-2B(z)}{2A(z)}\bB_{A}^{\pm}(z)\biggr|_{z=z(q)},
\end{align}
On the other hand, taking the derivative of (\ref{eq:rel2}) with respect to $q$ and using (\ref{eq:rel1}), we obtain
\begin{align}
\bU^{\pm\prime}(q)=-\bB_{A}^{\pm\prime}(z)+\frac{A'(z)}{2A(z)}\bB_{A}^{\pm}(z)\biggr|_{z=z(q)}.
\end{align}
Hence, we finally know the transformation rule of $\bC^{\pm}(z)$ under the transposition:
\begin{align}
(\bC^{\pm})^{\rmT}(z)=&\;\bC^{\pm}(z)-\bB_{A}^{\pm\prime}(z)+\frac{A'(z)-B(z)}{A(z)}\bB_{A}^{\pm}(z).
\label{eq:tC1}
\end{align}
In components, it reads as
\begin{align}
C_{21}^{\pm}(z)=C_{12}^{\pm}(z)-B_{A12}^{\pm\prime}(z)+\frac{A'(z)-B(z)}{A(z)}B_{A12}^{\pm}(z).
\label{eq:tC2}
\end{align}
No constraints on diagonal elements of $\bC^{\pm}$ follow from (\ref{eq:tC1}).

Instead of working with the gauge potential $\cW_{\cN}^{-}(z)$, it is often more convenient to use another function $Q(z)$ defined by
\begin{align}
\frac{\rmd\cW_{\cN}^{-}}{\rmd z}=\frac{1}{2A(z)}\left(\frac{N-1}{2}A'(z)-Q(z)\right),
\label{eq:WN-2}
\end{align}
or equivalently by \cite{GT04}
\begin{align}
Q(z)=B(z)+\frac{\cN-2}{2}A'(z).
\label{eq:defQ}
\end{align}
With this $Q(z)$, the covariant derivative (\ref{eq:covd1}), in particular, takes a simple form as
\begin{align}
\frac{D}{D z}=\frac{\rmd}{\rmd z}+\frac{Q(z)}{A(z)}.
\label{eq:covd2}
\end{align}

\section{$\fq(2)$ Lie-superalgebraic quasi-solvable models}
\label{sec:q2}

In~\cite{BK94}, a family of quasi-solvable matrix differential operators was constructed so that they preserve the two-component linear monomial space
\begin{align}
\tbcV_{\cN}^{-}=\begin{pmatrix} \tcV_{\cN}^{(\rmA)}\\ \tcV_{\cN}^{(\rmA)}\end{pmatrix},
\label{eq:VN1}
\end{align}
where $\tcV_{\cN}^{(\rmA)}$ is the so-called $\cN$-dimensional type A monomial subspace defined by
\begin{align}
\tcV_{\cN}^{(\rmA)}=\braket{1,z,\dots,z^{\cN-1}}.
\label{eq:Amon}
\end{align}
It was later found in [DJ01] that the Lie superalgebra which has the space (\ref{eq:VN1}) as a module is $\fq(2)$ which consists of 4 bosonic, denoted by $T_{\pm}$, $T_{0}$, $J$, and 4 fermionic, denoted by $Q_{\pm}$, $Q_{0}$, $\bar{Q}$, elements satisfying
\begin{align}
\begin{split}
\bigl[ T_{0}, T_{\pm}\bigr]=\pm T_{\pm},\qquad 
\bigl[ T_{+}, T_{-}\bigr]=-2T_{0},\\
\bigl[ Q_{\pm}, T_{0}\bigr]=\mp Q_{\pm},\quad
\bigl[ Q_{\pm}, T_{\mp}\bigr]=2Q_{0},\quad
\bigl[ Q_{0}, T_{\pm}\bigr]=-Q_{\pm},\\
\bigl\{ Q_{+}, Q_{-}\bigr\}=2J,\qquad
\bigl\{ Q_{\pm}, \bar{Q}\bigr\}=\mp T_{\pm},\\
\bigl\{ Q_{0}, Q_{0}\bigr\}=\bigl\{ \bar{Q}, \bar{Q}\bigr\}=J,\quad
\bigl\{ Q_{0}, \bar{Q}\bigr\}=T_{0},
\end{split}
\end{align}
where, as usual, $[A, B]=AB-BA$ and $\{A, B\}=AB+BA$. In Ref.~\cite{DJ01}, three inequivalent representations in terms of $2\times 2$ matrix linear differential operators were presented. In the present case of the linear space (\ref{eq:VN1}), the corresponding representation is the second one in \cite{DJ01} given by
\begin{align}
\begin{split}
\bJ=\begin{pmatrix} \frac{\cN}{2} & 0\\ 0 & \frac{\cN}{2}\end{pmatrix},\quad \bT_{+}=\begin{pmatrix} z^{2}\del_{z}-(\cN-1)z & 0\\ -1 & z^{2}\del_{z}-(\cN-1)z\end{pmatrix},\\
\bT_{0}=\begin{pmatrix} z\del_{z}-\frac{\cN}{2} & 0\\ 0 & z\del_{z}-\frac{\cN-2}{2}\end{pmatrix},\qquad
\bT_{-}=\begin{pmatrix} \del_{z} & 1\\ 0 & \del_{z}\end{pmatrix},\\
\bQ_{+}=\begin{pmatrix} 0 & 0\\ \sqrt{\cN} & 0\end{pmatrix},\quad
\bQ_{0}=-\frac{\sqrt{\cN}}{2}\begin{pmatrix} 1 & 0\\ 0 & -1\end{pmatrix},\quad
\bQ_{-}=\begin{pmatrix} 0 & \sqrt{\cN}\\ 0 & 0\end{pmatrix},\\
\bar{\bQ}=-\frac{1}{\sqrt{\cN}}\begin{pmatrix} z\del_{z}-\frac{\cN}{2} & z^{2}\del_{z}-(\cN-1)z\\ -\del_{z} & -z\del_{z}+\frac{\cN-2}{2}\end{pmatrix}.
\end{split}
\label{eq:repq}
\end{align}
To the best of our knowledge, quasi-solvable matrix models constructed from this representation have not been dully investigated so far. Before considering its universal enveloping algebra, we first note that in the representation (\ref{eq:repq}) the following identities hold:
\begin{align}
\begin{split}
\bT_{+}\bT_{-}=(\bT_{0})^{2}-\bT_{0}-\sqrt{\cN}\bar{\bQ}-\bJ^{2},\qquad
\bQ_{+}\bQ_{-}=\bJ+\sqrt{\cN}\bQ_{0},\\
\cN\bar{\bQ}=-\bT_{+}\bQ_{-}+2\bT_{0}\bQ_{0}+\bT_{-}\bQ_{+}-\sqrt{\cN},\quad
2\bQ_{\pm}\bQ_{0}=\mp\sqrt{\cN}\bQ_{\pm},
\end{split}
\end{align}
Owing to them together with anti-comutation relations, we need not take into account in the universal enveloping algebra a term containing $\bT_{+}\bT_{-}$, $\bar{\bQ}$, and any quadratic form of fermionic operators $\bQ_{i}\bQ_{j}$. For our present purpose, we also note that a term having $\bT_{i}\bT_{j}\bQ_{k}$ should be omitted, since such a term results in a second-order differential operator which is not proportional to $\bI_{2}$ and hence violates the constraint (\ref{eq:rel1}). To summarize, the most general $2\times 2$ matrix second-order linear differential operator constructed from the universal enveloping algebra of $\fq(2)$ which reduces to the form (\ref{eq:tHpm1}) with (\ref{eq:rel1}) and preserves the space (\ref{eq:VN1}) is expressed as
\begin{align}
\tbH^{\fq(2)}=&-\sum_{i,j=\pm,0}b_{ij}\bT_{i}\bT_{j}-\frac{1}{\sqrt{\cN}}\sum_{i=\pm,0; j=\pm}f_{ij}\bT_{i}\bQ_{j}
+\frac{2}{\sqrt{\cN}}\sum_{i=\pm,0}f_{i0}\bT_{i}\bQ_{0}\notag\\
&-\sum_{i=\pm,0}b_{i}\bT_{i}-\frac{1}{\sqrt{\cN}}\sum_{i=\pm}f_{i}\bQ_{i}+\frac{2}{\sqrt{\cN}}f_{0}\bQ_{0}-\frac{2}{\cN}b_{J}\bJ,
\label{eq:q2H1}
\end{align}
where $b_{ij}$, $f_{ij}$, $b_{i}$, $f_{i}$, and $b_{J}$ are all constants. The number of these parameters amounts to 21. Then, each element of $\tbH^{\fq(2)}$ has the form
\begin{align}
\Bigl( \tbH^{\fq(2)}\Bigr)_{ij} =-A_{ij}^{\fq(2)}(z)\del_{z}^{2}-B_{ij}^{\fq(2)}(z)\del_{z}-C_{ij}^{\fq(2)}(z),
\end{align}
where $A_{ij}^{\fq(2)}(z)=A^{\fq(2)}(z)\delta_{ij}$ is given by
\begin{align}
A^{\fq(2)}(z)=b_{++}z^{4}+b_{+0}z^{3}+b_{00}z^{2}+b_{0-}z+b_{--}.
\end{align}
The first- and zeroth-order coefficients in the $(1,1)$-component are
\begin{align}
B_{11}^{\fq(2)}(z)=&-2(\cN-2)b_{++}z^{3}-\left( \frac{3\cN-4}{2}b_{+0}-f_{+0}-b_{+}\right) z^{2}\notag\\
&-\left[ (\cN-1)b_{00}-f_{00}-b_{0}\right] z-\frac{\cN}{2}b_{0-}+f_{-0}+b_{-},\\
C_{11}^{\fq(2)}(z)=&\;(\cN-1)(\cN-2)b_{++}z^{2}+(\cN-1)\left( \frac{\cN}{2}b_{+0}-f_{+0}-b_{+}\right) z\notag\\
&+\frac{\cN^{2}}{4}b_{00}-\frac{\cN}{2}(f_{00}+b_{0})+f_{-+}+f_{0}+b_{J},
\end{align}
in the $(1,2)$-component are
\begin{align}
B_{12}^{\fq(2)}(z)&=f_{+-}z^{2}+(b_{0-}+f_{0-})z+2b_{--}+f_{--},\\
C_{12}^{\fq(2)}(z)&=-(\cN-1)f_{+-}z-\frac{\cN}{2}(b_{0-}+f_{0-})-f_{-0}+b_{-}+f_{-},
\end{align}
in the $(2,1)$-component are
\begin{align}
B_{21}^{\fq(2)}(z)&=-(2b_{++}-f_{++})z^{2}-(b_{+0}-f_{0+})z+f_{-+},\\
C_{21}^{\fq(2)}(z)&=(\cN-1)(2b_{++}-f_{++})z+\frac{\cN}{2}b_{+0}-\frac{\cN-2}{2}f_{0+}-f_{+0}-b_{+}+f_{+},
\end{align}
and in the $(2,2)$-component are
\begin{align}
B_{22}^{\fq(2)}(z)=&-2(\cN-2)b_{++}z^{3}-\left[ \frac{3(\cN-2)}{2}b_{+0}+f_{+0}-b_{+}\right] z^{2}\notag\\
&-\left[ (\cN-3)b_{00}+f_{00}-b_{0}\right] z-\frac{\cN-2}{2}b_{0-}-f_{-0}+b_{-},\\
C_{22}^{\fq(2)}(z)=&\;(\cN-1)(\cN-2)b_{++}z^{2}+(\cN-1)\left( \frac{\cN-2}{2}b_{+0}+f_{+0}-b_{+}\right) z\notag\\
&+\frac{(\cN-2)^{2}}{4}b_{00}+\frac{\cN-2}{2}(f_{00}-b_{0})-f_{+-}-f_{0}+b_{J}.
\end{align}
It follows from the representation (\ref{eq:repq}) that except for $\bT_{+}$ and $\bar{\bQ}$ all the generators preserve the linear space (\ref{eq:VN1}) for all $\cN\in\bbN$. Hence, the system $\bH^{-}$ preserves an infinite flag consist of the finite-dimensional linear spaces (\ref{eq:VN1})
\begin{align}
\tbcV_{1}^{-}\subset\tbcV_{2}^{-}\subset\dots\subset\tbcV_{\cN}^{-}\cdots,
\label{eq:inflag}
\end{align}
and thus is solvable if $b_{++}=b_{+0}=f_{+i}=b_{+}=0$ ($i=\pm,0$).

\section{$\cN$-fold SUSY Approach to the $\fq(2)$ Case}
\label{sec:Nf1}

To construct an $\cN$-fold SUSY system from the linear space (\ref{eq:VN1}), we first prepare a gauged $\cN$-fold supercharge component $\tbP_{\cN}^{-}$ such that it annihilates any elements of the linear space. Such an operator of order $\cN$ is uniquely determined as
\begin{align}
\tbP_{\cN}^{-}=\begin{pmatrix} \tP_{\cN}^{(\rmA)} & 0\\ 0 & \tP_{\cN}^{(\rmA)}\end{pmatrix}=(z')^{\cN}\begin{pmatrix} \del_{z}^{\cN} & 0\\ 0 & \del_{z}^{\cN}\end{pmatrix},\qquad \tbP_{\cN}^{+}=(-z')^{\cN}\begin{pmatrix} D_{z}^{\cN} & 0\\ 0 & D_{z}^{\cN}\end{pmatrix},
\label{eq:tbP1}
\end{align}
where $\tP_{\cN}^{(\rmA)}=(z')^{\cN}\del_{z}^{\cN}$ is a gauged type A $\cN$-fold supercharge component annihilating an $\cN$-dimensional type A monomial subspace $\tcV_{\cN}^{(\rmA)}$, and $D_{z}=D/Dz$ is the abbreviation for the covariant derivative. If we denote the $(i,j)$th component of $\tbH^{\pm}$ as $\tH_{ij}^{\pm}=(\tbH^{\pm})_{ij}$, the gauged version of the intertwining relation (\ref{eq:Nf1}) is represented in terms of components as
\begin{align}
\tbP_{\cN}^{-}\tbH^{-}-\tbH^{+}\tbP_{\cN}^{-}=0\quad\Llra\quad \tP_{\cN}^{(\rmA)}\tH_{ij}^{-}-\tH_{ij}^{+}\tP_{\cN}^{(\rmA)}=0,
\end{align}
for all $i,j=1,2$. The latter relation just means that each $\tH_{ij}^{\pm}$ constitutes a pair of type A $\cN$-fold SUSY. Noting that $A_{ij}(z)=A(z)\delta_{ij}$ with $A(z)\neq 0$ from (\ref{eq:rel1}), we can employ the known result of \cite{Ta03a} to obtain
\begin{align}
\tH_{ij}^{\pm}=&-A(z)\delta_{ij}\frac{\rmd^{2}}{\rmd z^{2}}-\left( Q_{ij}(z)-\frac{\cN-2}{2}A'(z)\delta_{ij}\right)\frac{\rmd}{\rmd z}-\frac{(\cN-1)(\cN-2)}{12}A''(z)\delta_{ij}\notag\\
&+\frac{\cN-1}{2}Q'_{ij}(z)-R_{ij}-\frac{(1\pm 1)\cN}{2}\left( Q'_{ij}(z)-\frac{A'(z)Q_{ij}(z)}{2A(z)}\right),
\label{eq:tHij1}
\end{align}
where $R_{ij}$ are constants, and $A(z)$ for $\cN=1,2$ and $Q_{ij}(z)$ for $\cN=1$ are arbitrary functions of $z$ but must otherwise have the polynomial form of
\begin{align}
A(z)&=a_{4}z^{4}+a_{3}z^{3}+a_{2}z^{2}+a_{1}z+a_{0}\quad\text{for}\quad\cN\geq 3,\\
Q_{ij}(z)&=b_{ij,2}z^{2}+b_{ij,1}z+b_{ij,0}\quad\text{for}\quad\cN\geq 2,
\label{eq:Qij1}
\end{align}
with $a_{k}$ ($k=0,\dots,4$) and $b_{ij,k}$ ($k=0,1,2$) being constants. Hence, there are 21 independent parameters in total for $\cN\geq 3$, and 16 for $\cN=2$. Comparing now the obtained system $\tbH^{-}$, which satisfies only the intertwining (\ref{eq:Nf1}) with respect to $\tbP_{\cN}^{-}$ in (\ref{eq:tbP1}) and thus preserves the space (\ref{eq:VN1}), to the $\fq(2)$ quasi-solvable matrix Hamiltonian $\tbH^{\fq(2)}$ given by (\ref{eq:q2H1}), we see that they are completely identical with each other. In particular, for $\cN\geq 3$, they are connected by the following correspondence of the 21 parameters:
\begin{align}
\begin{split}
&a_{4}=b_{++},\quad a_{3}=b_{+0},\quad a_{2}=b_{00},\quad a_{1}=b_{0-},\quad a_{0}=b_{--},\\
&b_{11,2}=-b_{+0}+f_{+0}+b_{+},\quad b_{11,1}=-b_{00}+f_{00}+b_{0},\quad b_{11,0}=-b_{0-}+f_{-0}+b_{-},\\
&b_{12,2}=f_{+-},\quad b_{12,1}=b_{0-}+f_{0-},\quad b_{12,0}=2b_{--}+f_{--},\quad
b_{21,2}=-2b_{++}+f_{++},\\
&b_{21,1}=-b_{+0}+f_{0+},\quad b_{21,0}=f_{-+},\quad
b_{22,2}=-f_{+0}+b_{+},\quad b_{22,1} \lra b_{00}-f_{00}+b_{0},\\
&b_{22,0}=-f_{-0}+b_{-},\quad
R_{11}=-\frac{5\cN^{2}-12\cN+10}{12}b_{00}+\frac{f_{00}+b_{0}}{2}-f_{-+}-f_{0}-b_{J},\\
&R_{12}=\frac{b_{0-}+f_{0-}}{2}+f_{-0}-b_{-}-f_{-},\quad 
R_{21}=-\frac{b_{+0}+f_{0+}}{2}+f_{+0}+b_{+}-f_{+},\\
&R_{22}=-\frac{5\cN^{2}-12\cN+10}{12}b_{00}+\frac{f_{00}-b_{0}}{2}+f_{+-}+f_{0}-b_{J}.
\end{split}
\end{align}
Hence, the intertwining relation (\ref{eq:Nf1}) with respect to $\tbP_{\cN}^{-}$ in (\ref{eq:tbP1}) can solely reproduce the $\fq(2)$ Lie-superalgebraic quasi-solvable matrix operator. An advantage of the intertwining approach is that one can simultaneously construct another quasi-solvable operator $\tbH^{+}$ which is almost isospectral to $\tbH^{-}$. However, it should be reminded that it is in general quite difficult to solve an intertwining relation for an arbitrary natural number $\cN$.

We shall next see what constraints emerge when the transposition symmetry is imposed on the system. Applying first the condition (\ref{eq:tB2}) for $\bB$ with (\ref{eq:defQ}), we have
\begin{align}
\begin{split}
Q(z)&=Q_{11}(z)=Q_{22}(z)\left[ =b_{2}z^{2}+b_{1}z+b_{0}\quad\text{for}\quad \cN\geq 2\right],\\
Q_{A}(z)&=Q_{12}(z)=-Q_{21}(z)\left[ =b_{A2}z^{2}+b_{A1}z+b_{A0}\quad\text{for}\quad \cN\geq 2\right].
\end{split}
\label{eq:QQA}
\end{align}
The condition (\ref{eq:tC2}) for $\bC$ then reads as
\begin{align*}
(\cN-2)Q'_{A}+\frac{\cN A'-2Q}{2A}Q_{A}+R_{21}-R_{12}=Q'_{A}-\frac{A'Q_{A}}{2A}=0,
\end{align*}
which leads for an arbitrary $\cN$ to the following solution of a non-trivial $Q_{A}(z)$:
\begin{align}
A(z)=c_{A}Q_{A}(z)^{2},\qquad Q(z)=c_{A}Q_{A}(z)\left[ 2(\cN-1)Q'_{A}(z)+R_{21}-R_{12}\right],
\label{eq:AQ1}
\end{align}
where $c_{A}$ is an integral constant, or to the trivial solution:
\begin{align}
Q_{A}(z)=0,\qquad R_{21}=R_{12}.
\label{eq:AQ2}
\end{align}
For the case $\cN\geq 2$ where both $Q(z)$ and $Q_{A}(z)$ must be polynomials of at most second order, as indicated in (\ref{eq:QQA}), the non-trivial solution (\ref{eq:AQ1}) is compatible only if $b_{A2}=0$ and thus
\begin{align}
A(z)=(b_{A1}z+b_{A0})^{2},\quad Q(z)=c_{A}[2(\cN-1)b_{A1}+R_{21}-R_{12}](b_{A1}z+b_{A0}).
\end{align}
Hence, for the non-trivial solution, $A(z)$ must be a polynomial of at most second order while $Q(z)$ and $Q_{A}(z)$ of at most first order. In this case, each $\tbH^{\pm}$ preserves an infinite flag of finite-dimensional two-component linear space and turns to be solvable.

At last, we shall impose the second algebraic relation (\ref{eq:Nf2}) for $\cN$-fold SUSY. Using (\ref{eq:tbP1}), we obtain
\begin{align}
\tbP_{\cN}^{-}\tbP_{\cN}^{+}=(-1)^{\cN}(z')^{2\cN}( \del_{\cN, z})^{\cN}D_{z}^{\cN}\bI_{2},\\
\tbP_{\cN}^{+}\tbP_{\cN}^{-}=(-1)^{\cN}(z')^{2\cN}(D_{\cN, z})^{\cN}\del_{z}^{\cN}\bI_{2},
\end{align}
where
\begin{align}
\del_{\cN, z}=\del_{z}+\frac{\cN A'}{2A},\qquad D_{\cN, z}=D_{z}+\frac{\cN A'}{2A}=\del_{z}+\frac{2Q+\cN A'}{2A}.
\label{eq:dNz}
\end{align}
Hence, the right hand side of (\ref{eq:Nf2}) must be proportional to $\bI_{2}$ as well. Given that the diagonal elements of $\bH^{\pm}$ are all non-trivial, it is possible only if $\bH^{\pm}+\bC_{0}$ and $\bC_{k}$ ($k=1,\dots,\cN-2$) are all proportional to $\bI_{2}$, too. From (\ref{eq:tHij1}), they are all satisfied only if, for all $i\neq j$,
\begin{align}
Q_{ij}(z)=(\bC_{0})_{ij}-R_{ij}=(\bC_{k})_{ij}=0\quad (k=1,\dots,\cN-1),
\end{align}
which can be realized only for the trivial solution (\ref{eq:AQ2}) of the transposition symmetry. It means that the algebraic condition (\ref{eq:Nf2}) is stronger than the intertwining relation (\ref{eq:Nf1}) supplemented with the transposition symmetry. This example illustrates a violation of equivalence between quasi-solvability and $\cN$-fold SUSY in matrix models, in contrast to the scalar case. To summarize the results in this section, we have found that the most general $\cN$-fold SUSY system constructed from the two-component linear space (\ref{eq:VN1}) is essentially equivalent to the scalar type A $\cN$-fold SUSY.

\section{$\fosp(2/2)$ Lie-superalgebraic quasi-solvable models}
\label{sec:osp22}

We shall next consider another two-component monomial subspace given by
\begin{align}
\tbcV_{\cN}^{-}=\begin{pmatrix} \tcV_{\cN-1}^{(\rmA)}\\ \tcV_{\cN}^{(\rmA)}\end{pmatrix},
\label{eq:VN2}
\end{align}
where $\tcV_{\cN}^{(\rmA)}$ is again an $\cN$-dimensional type A monomial subspace (\ref{eq:Amon}). Quasi-solvable matrix operators preserving this space have been best studied in the literature \cite{Tu92b}, and the Lie superalgebra which has the linear space (\ref{eq:VN2}) as a module is $\fosp(2/2)$. It consists of 4 bosonic, denoted by $T_{\pm}$, $T_{0}$, $J$, and 4 fermionic, denoted by $Q_{\pm}$, $\bar{Q}_{\pm}$, elements, as in the previous $\fq(2)$, but they satisfy the following different commutation and anti-commutation relations:
\begin{align}
\begin{split}
\bigl[ T_{0}, T_{\pm}\bigr]=\pm T_{\pm},\qquad 
\bigl[ T_{+}, T_{-}\bigr]=-2T_{0},\\
\bigl[ \bar{Q}_{\pm}, T_{0}\bigr]=\mp\frac{1}{2}\bar{Q}_{\pm},\quad 
\bigl[ Q_{\pm}, J\bigr]=-\frac{1}{2}Q_{\pm},\quad 
\bigl[ \bar{Q}_{\pm}, J\bigr]=\frac{1}{2}\bar{Q}_{\pm},\\
\bigl[ Q_{\mp}, T_{\pm}\bigr]=\pm Q_{\pm},\quad 
\bigl[ \bar{Q}_{\mp}, T_{\pm}\bigr]=\mp\bar{Q}_{\pm},\quad
\bigl[ Q_{\pm}, T_{0}\bigr]=\mp\frac{1}{2}Q_{\pm},\\
\bigl\{ \bar{Q}_{\pm}, Q_{\pm}\bigr\}=\pm T_{\pm},\qquad
\bigl\{ \bar{Q}_{\pm}, Q_{\mp}\bigr\}=J\pm T_{0},
\end{split}
\end{align}
A representation in terms of $2\times 2$ matrix linear differential operators is given by
\begin{align}
\begin{split}
\bJ=\begin{pmatrix} -\frac{\cN}{2} & 0\\ 0 & -\frac{\cN-1}{2}\end{pmatrix},\quad 
\bT_{+}=\begin{pmatrix} z^{2}\del_{z}-(\cN-2)z & 0\\ 0 & z^{2}\del_{z}-(\cN-1)z \end{pmatrix},\\
\bT_{0}=\begin{pmatrix} z\del_{z}-\frac{\cN-2}{2} & 0\\ 0 & z\del_{z}-\frac{\cN-1}{2}\end{pmatrix},\quad 
\bT_{-}=\begin{pmatrix} \del_{z} & 0\\ 0 & \del_{z}\end{pmatrix},\quad 
\bQ_{+}=\begin{pmatrix} 0 & 0\\ z & 0\end{pmatrix},\\
\bQ_{-}=\begin{pmatrix} 0 & 0\\ 1 & 0\end{pmatrix},\qquad 
\bar{\bQ}_{+}=\begin{pmatrix} 0 & z\del_{z}-(\cN-1)\\ 0 & 0\end{pmatrix},\qquad 
\bar{\bQ}_{-}=\begin{pmatrix} 0 & -\del_{z}\\ 0 & 0\end{pmatrix}.
\end{split}
\label{eq:reposp}
\end{align}
It is evident that any operator composed from its universal enveloping algebra preserves the linear space (\ref{eq:VN2}) and is thus quasi-solvable. The most general $2\times 2$ matrix linear differential operator of (at most) second order which is available in this way was discussed in \cite{Tu92b}. In our case, we must take into account the fact that the coefficients of second order in diagonal components are identical as in (\ref{eq:rel1}) and that the off-diagonal components are at most first order. For $\cN\geq 4$, we find that the most general operator which meets our purpose is expressed in terms of the $\fosp(2/2)$ generators as
\begin{align}
\tbH^{\fosp}=&-\sum_{i,j=\pm,0}b_{ij}\bT_{i}\bT_{j}-\sum_{i=\pm,0}b_{iJ}\bT_{i}\bJ-\sum_{i=\pm}f_{+i}\bT_{+}\bQ_{i}-\sum_{i=0,-}f_{i-}\bT_{i}\bQ_{-}\notag\\
&-\sum_{i=\pm,0}b_{i}\bT_{i}-b_{J}\bJ-\sum_{i=\pm}f_{i}\bQ_{i}-\sum_{i=\pm}\bar{f}_{i}\bar{\bQ}_{i}-b_{I}\bI_{2},
\label{eq:Hosp}
\end{align}
where $b_{\ldots}$, $f_{\ldots}$, and $\bar{f}_{\ldots}$ are all constants. Then, each element of $\tbH^{\fosp}$ has the form
\begin{align}
\left( \tbH^{\fosp}\right)_{ij} =-A_{ij}^{\fosp}(z)\del_{z}^{2}-B_{ij}^{\fosp}(z)\del_{z}-C_{ij}^{\fosp}(z),
\end{align}
where the second-order coefficients $A_{ij}^{\fosp}(z)=A^{\fosp}(z)\delta_{ij}$ are given by
\begin{align}
A^{\fosp}(z)=b_{++}z^{4}+b_{+0}z^{3}+b_{+-}z^{2}+b_{0-}z+b_{--}.
\label{eq:Aosp}
\end{align}
The coefficients of first and zeroth order are respectively given for the $(1,1)$ component by
\begin{align}
B_{11}^{\fosp}(z)=&-2(\cN-3)b_{++}z^{3}-\left( \frac{3\cN-8}{2}b_{+0}+\frac{\cN}{2}b_{+J}-b_{+}\right) z^{2}\notag\\
&-\left[ (\cN-2)b_{+-}+\frac{\cN}{2}b_{0J}-b_{0}\right] z-\frac{\cN-2}{2}b_{0-}-\frac{\cN}{2}b_{-J}+b_{-},
\label{eq:Bosp11}\\
C_{11}^{\fosp}(z)=&\;(\cN-2)(\cN-3)b_{++}z^{2}+(\cN-2)\left[ \frac{\cN-2}{2}b_{+0}+\frac{\cN}{2}b_{+J}-b_{+}\right] z\notag\\
&+\frac{\cN(\cN-2)}{4}b_{0J}-\frac{\cN-2}{2}b_{0}-\frac{\cN}{2}b_{J}+b_{I},
\label{eq:Cosp11}
\end{align}
the $(1,2)$ component by
\begin{align}
B_{12}^{\fosp}(z)=\bar{f}_{+}z-\bar{f}_{-},\qquad
C_{12}^{\fosp}(z)=-(\cN-1)\bar{f}_{+},
\label{eq:BCosp12}
\end{align}
the $(2,1)$ component by
\begin{align}
B_{21}^{\fosp}(z)=&f_{++}z^{3}+f_{+-}z^{2}+f_{0-}z+f_{--},\\
C_{21}^{\fosp}(z)=&-(\cN-2)f_{++}z^{2}-\left[ (\cN-1)f_{+-}-f_{+}\right] z-\frac{\cN-1}{2}f_{0-}+f_{-},
\end{align}
and the $(2,2)$ component by
\begin{align}
B_{22}^{\fosp}(z)=&-2(\cN-2)b_{++}z^{3}-\left( \frac{3\cN-5}{2}b_{+0}+\frac{\cN-1}{2}b_{+J}-b_{+}\right) z^{2}\notag\\
&-\left[ (\cN-1)b_{+-}+\frac{\cN-1}{2}b_{0J}-b_{0}\right] z-\frac{\cN-1}{2}(b_{0-}+b_{-J})+b_{-},\\
C_{22}^{\fosp}(z)=&\;(\cN-1)(\cN-2)b_{++}z^{2}+(\cN-1)\left[ \frac{\cN-1}{2}(b_{+0}+b_{+J})-b_{+}\right] z\notag\\
&+\frac{(\cN-1)^{2}}{4}b_{0J}-\frac{\cN-1}{2}(b_{0}+b_{J})+b_{I}.
\label{eq:Cosp22}
\end{align}
In the representation (\ref{eq:reposp}) of $\fosp(2/2)$, that except for $\bT_{+}$ and $\bar{\bQ}$ all the generators preserve the linear space (\ref{eq:VN2}) for all $\cN\in\bbN$. Hence, the system $\bH^{-}$ preserves an infinite flag (\ref{eq:inflag}) ordered by the linear spaces (\ref{eq:VN2}) of increasing dimension, and thus is solvable if $b_{++}=b_{+0}=f_{+i}=b_{+}=0$ ($i=\pm,0$).

For $\cN=3$, the upper component of the linear space (\ref{eq:VN2}) is $\tcV_{2}^{(\rmA)}=\langle 1,z\rangle$ and is trivially preserved by $\del_{z}^{2}$. Hence, the $(1,1)$-component of $\tbH^{\fosp}$ can have any coefficient of $\del_{z}^{2}$. However, the constraint (\ref{eq:rel1}) requires that it must be identical to that of the $(2,2)$-component. Thus, $\tbH^{\fosp}$ in the $\cN=3$ case is also given by the same as in the above $\cN\geq 4$ case.

For $\cN=2$, the upper component of the linear space (\ref{eq:VN2}) is $\tcV_{1}^{(\rmA)}=\langle 1\rangle$ and is trivially preserved by $\del_{z}^{2}$ and $\del_{z}$. Hence, $(1,1)$- and $(2,1)$-component of $\tbH^{\fosp}$ admit any coefficients except for in the zeroth order. On the other hand, the lower component of the space is $\tcV_{2}^{(\rmA)}=\langle 1,z\rangle$ and is trivially preserved by $\del_{z}^{2}$. Hence, the $(2,2)$-component of $\tbH^{\fosp}$ can have any coefficient of $\del_{z}^{2}$. Considering the both components, we see that the second-order coefficient $A(z)$, which is common in the $(1,1)$- and $(2,2)$-component, can be arbitrary and is free from the algebraic structure (\ref{eq:Hosp}). Furthermore, the first-order coefficients in the $(1,1)$- and $(2,1)$-component can be also arbitrary. Hence, we can set the five parameters $\{ b_{ij}\}$ in (\ref{eq:Aosp}), $\{ f_{+i}\}$, and $\{ f_{i-}\}$ to be zero, and replace the second-order coefficients in the $(1,1)$- and $(2,2)$-component by an arbitrary function $A(z)$, and add other arbitrary functions $B_{11}^{ad}(z)$ and $B_{21}^{ad}(z)$ to the first-order coefficients in the $(11)$- and $(2,1)$-component, respectively. Hence, for the $\cN=2$ case we have, instead of (\ref{eq:Bosp11})--(\ref{eq:Cosp22}),
\begin{align}
\begin{split}
B_{11}^{\fosp}(z)&=B_{11}^{ad}(z)+(b_{+}-b_{+J})z^{2}+(b_{0}-b_{0J})z+b_{-}-b_{-J},\quad
C_{11}^{\fosp}(z)=b_{I}-b_{J},\\
B_{12}^{\fosp}(z)&=\bar{f}_{+}z-\bar{f}_{-},\quad C_{12}^{\fosp}(z)=-\bar{f}_{+},\quad 
B_{21}^{\fosp}(z)=B_{21}^{ad}(z),\quad C_{21}^{\fosp}(z)=f_{+}z+f_{-},\\
B_{22}^{\fosp}(z)&=\left( b_{+}-\frac{b_{+J}}{2}\right)z^{2}+\left( b_{0}-\frac{b_{0J}}{2}\right)z+b_{-}-\frac{b_{-J}}{2},\\
C_{22}^{\fosp}(z)&=\left( \frac{b_{+J}}{2}-b_{+}\right)z+\frac{b_{0J}}{4}-\frac{b_{0}+b_{J}}{2}+b_{I}.
\end{split}
\end{align}
In the above, we can regard each of the two combinations $b_{i}-b_{iJ}$ in $B_{11}^{\fosp}(z)$ and $b_{i}-b_{iJ}/2$ in $B_{22}^{\fosp}(z)$ and $C_{22}^{\fosp}(z)$ for $i=\pm,0$ as independent parameters so that the polynomial parts in $B_{11}^{\fosp}(z)$ can be included in the arbitrary function $B_{11}^{ad}(z)$. It is formally achieved by setting $b_{iJ}=b_{i}$ ($i=\pm,0$) to yield
\begin{align}
\begin{split}
B_{11}^{\fosp}(z)&=B_{11}^{ad}(z),\qquad C_{11}^{\fosp}(z)=b_{I}-b_{J},\qquad 
B_{12}^{\fosp}(z)=\bar{f}_{+}z-\bar{f}_{-},\\
C_{12}^{\fosp}(z)&=-\bar{f}_{+},\qquad B_{21}^{\fosp}(z)=B_{21}^{ad}(z),\qquad 
C_{21}^{\fosp}(z)=f_{+}z+f_{-},\\
B_{22}^{\fosp}(z)&=\frac{b_{+}}{2}z^{2}+\frac{b_{0}}{2}z+\frac{b_{-}}{2},\qquad 
C_{22}^{\fosp}(z)=-\frac{b_{+}}{2}z-\frac{b_{0}}{4}-\frac{b_{J}}{2}+b_{I}.
\end{split}
\label{eq:ospN2}
\end{align}
Hence, the quasi-solvable matrix model in the $\cN=2$ case is characterized by 3 arbitrary functions $A(z)$, $B_{11}^{ad}(z)$, $B_{21}^{ad}(z)$, and 9 free parameters $b_{i}$ (i=$\pm,0$), $b_{I}$, $b_{J}$, $f_{\pm}$, and $\bar{f}_{\pm}$.

For $\cN=1$, the upper component of the linear space (\ref{eq:VN2}) is trivial,  $\tcV_{0}^{(\rmA)}=\emptyset$, hence $(1,1)$- and $(2,1)$-component of $\tbH^{\fosp}$ can be arbitrary. On the other hand, the lower component of the space is $\tcV_{1}^{(\rmA)}=\langle 1\rangle$ and is trivially preserved by $\del_{z}^{2}$ and $\del_{z}$. Hence, $(1,2)$- and $(2,2)$-component of $\tbH^{\fosp}$ admit any coefficients except for in the zeroth order. The $(1,2)$-component raises the lower component of $\tbcV_{1}^{-}$ to the upper one which must be trivial. On the other hand, the $(2,2)$-component maps the lower component to itself which must be a constant. To summarize the analysis, therefore, $\tbH^{\fosp}$ is essentially free from the algebraic structure (\ref{eq:Hosp}) and the only residing constraint is
\begin{align}
C_{12}^{\fosp}(z)=0,\qquad C_{22}^{\fosp}(z)=c_{0},
\label{eq:ospN1}
\end{align}
where $c_{0}$ is a constant. Hence, the quasi-solvable matrix model in the $\cN=1$ case consists of 7 arbitrary functions $A^{\fosp}(z)$, $B_{ij}^{\fosp}(z)$, $C_{1i}^{\fosp}(z)$ ($i,j=1,2$) and 1 arbitrary constant $c_{0}$.

\section{$\cN$-fold SUSY Approach to the $\fosp(2/2)$ Case}
\label{sec:Nf2}

To construct an $\cN$-fold SUSY system from the linear space (\ref{eq:VN2}), we shall begin with a gauged $\cN$-fold supercharge component $\tbP_{\cN}^{-}$ which annihilates any elements of the linear space. The most general matrix linear diferential operator of $\cN$th order which matches the form of (\ref{eq:Pm1}) is
\begin{align}
\begin{split}
\tbP_{\cN}^{-}&=\begin{pmatrix} \tP_{\cN}^{(\rmA)}+\beta z'\tP_{\cN-1}^{(\rmA)} & 0\\ \alpha z'\tP_{\cN-1}^{(\rmA)} & \tP_{\cN}^{(\rmA)}\end{pmatrix}=(z')^{\cN}\begin{pmatrix} \del_{z}^{\cN}+\beta\del_{z}^{\cN-1} & 0\\ \alpha\del_{z}^{\cN-1} & \del_{z}^{\cN}\end{pmatrix},\\
\tbP_{\cN}^{+}&=(-z')^{\cN}\begin{pmatrix} D_{z}^{\cN}-D_{z}^{\cN-1}\beta & -D_{z}^{\cN-1}\alpha\\ 0 & D_{z}^{\cN}\end{pmatrix},
\end{split}
\label{eq:tbP2}
\end{align}
where $\tP_{\cN}^{(\rmA)}$ is again a gauged type A $\cN$-fold supercharge component.
In contrast to the $\fq(2)$ case in Section~\ref{sec:Nf2}, it is not unique and admits two (at present) arbitrary functions $\alpha(z)$ and $\beta(z)$ in the expression (\ref{eq:tbP2}). This ambiguity is related to the fact that the dimension of the upper linear space in (\ref{eq:VN2}) is less than $\cN$. Moreover, as we shall show below, the operator $\tbP_{\cN}^{-}$ in (\ref{eq:tbP2}) cannot characterize the space completely due to the latter fact.

To begin with, let us consider a diagonal $\tbP_{\cN}^{-}$ case realized by putting $\alpha(z)=0$. In this particular case, the $\cN$th-order operator appeared in the $(1,1)$-component of $\tbP_{\cN}^{-}$ for $\beta(z)\neq 0$ can be written as
\begin{align}
\tP_{\cN}^{(\rmA)}+\beta z'\tP_{\cN-1}^{(\rmA)}=(z')^{\cN}(\del_{z}+\beta)\del_{z}^{\cN-1}=\rme^{\cW_{\cN}^{-}}P_{\cN}^{(\rmB)-}\rme^{-\cW_{\cN}^{-}},
\end{align}
where the gauge factor $\cW_{\cN}^{-}$ is the same as given by (\ref{eq:WN-1}) or (\ref{eq:WN-2}), and $P_{\cN}^{(\rmB)-}$ is a so-called type B $\cN$-fold supercharge component defined by~\cite{GT04}
\begin{align}
P_{\cN}^{(\rmB)-}=&\left( \frac{\rmd}{\rmd q}+W(q)-F(q)-\frac{\cN-1}{2}E(q)\right)\notag\\
&\times\prod_{k=0}^{\cN-2}\left( \frac{\rmd}{\rmd q}+W(q)+\frac{\cN-1-2k}{2}E(q)\right).
\label{eq:BPN-}
\end{align}
In our present case, the three functions $E(q)$, $W(q)$, and $F(q)$ in (\ref{eq:BPN-}) are expressed in terms of the already introduced functions as
\begin{align}
E(q)=\frac{z''(q)}{z'(q)},\qquad W(q)=-\frac{Q(z(q))}{z'(q)},\qquad F(q)=-\beta(z(q))z'(q).
\end{align}
Hence, the $(1,1)$-component of $\tbP_{\cN}^{-}$ is nothing but a gauged type B $\cN$-fold supercharge component:
\begin{align}
\tP_{\cN}^{(\rmA)}+\beta z'\tP_{\cN-1}^{(\rmA)}=\tP_{\cN}^{(\rmB)}.
\end{align}
Then, the gauged version of the intertwining relation (\ref{eq:Nf1}) consists of the following components:
\begin{align}
\begin{split}
\tP_{\cN}^{(\rmB)}\tH_{11}^{-}-\tH_{11}^{+}\tP_{\cN}^{(\rmB)}=0,\qquad
\tP_{\cN}^{(\rmB)}\tH_{12}^{-}-\tH_{12}^{+}\tP_{\cN}^{(\rmA)}=0,\\
\tP_{\cN}^{(\rmA)}\tH_{21}^{-}-\tH_{21}^{+}\tP_{\cN}^{(\rmB)}=0,\qquad
\tP_{\cN}^{(\rmA)}\tH_{22}^{-}-\tH_{22}^{+}\tP_{\cN}^{(\rmA)}=0.
\end{split}
\label{eq:intB}
\end{align}
The first relation in (\ref{eq:intB}) just means that $\tH_{11}^{\pm}$ forms a pair of type B $\cN$-fold SUSY for $\beta\neq 0$. Unfortunately, the most general form of type B $\cN$-fold SUSY for an arbitrary $\cN\geq 4$ has not been known yet, and only its particular subclass has been revealed~\cite{GT04}. Thus, at present we shall content ourselves with employing the result of the latter particular case. It is characterized by the choice $F(z(q))=z'(q)/z(q)$, which corresponds in our case to $\beta(z)=-1/z$. According to~\cite{GT04,GT05}, the most general gauged type B $\cN$-fold SUSY Hamiltonian of the minus component for an arbitrary integral $\cN\geq 3$ which should correspond to our $\tH_{11}^{-}$ has the following coefficients:
\begin{align}
A(z)&=a_{4}z^{4}+a_{3}z^{3}+a_{2}z^{2}+a_{1}z+a_{0},
\label{eq:A2}\\
B_{11}^{-}(z)&=-2(\cN-2)a_{4}z^{3}-(2\cN-3)a_{3}z^{2}+\left[ b_{11,1}-(\cN-2)a_{2}\right] z-(\cN-1)a_{1},
\label{eq:B11-2}\\
C_{11}^{-}(z)&=\cN(\cN-3)a_{4}z^{2}+\cN(\cN-2)a_{3}z+c_{11,0},
\label{eq:C11-2}
\end{align}
where $a_{i}$ ($i=0,\dots,4$), $b_{11,1}$, and $c_{11,0}$ are constants. Similarly, the fourth relation in (\ref{eq:intB}) just says that $\tH_{22}^{\pm}$ forms a pair of type A $\cN$-fold SUSY pair. Hence, $\tH_{22}^{\pm}$ is entirely the same as the one given in (\ref{eq:tHij1})--(\ref{eq:Qij1}). Explicitly for $\cN\geq 3$, $A(z)$ is the same as (\ref{eq:A2}) while the coefficients in $\tH_{22}^{-}$ read as
\begin{align}
B_{22}^{-}(z)=&-2(\cN-2)a_{4}z^{3}+\left[ b_{22,2}-\frac{3(\cN-2)}{2}a_{3}\right] z^{2}\notag\\
&+\left[ b_{22,1}-(\cN-2)a_{2}\right] z+b_{22,0}-\frac{\cN-2}{2}a_{1},
\label{eq:B22-2}\\
C_{22}^{-}(z)=&(\cN-1)(\cN-2)a_{4}z^{2}+(\cN-1)\left( \frac{\cN-2}{2}a_{3}-b_{22,2}\right) z+c_{22,0},
\label{eq:C22-2}
\end{align}
where $b_{22,i}$ ($i=0,1,2$) and $c_{22,0}$ are constants. On the other hand, the second and third relations in (\ref{eq:intB}) have peculiar forms and their general solutions for an arbitrary $\cN\in\bbN$ have not been known to us. However, it is clear that the product operators $\tbP_{\cN}^{\mp}\tbP_{\cN}^{\pm}$ are diagonal for $\beta(z)=0$ and that the algebraic constraint (\ref{eq:Nf2}) requires the trivial solutions $H_{12}^{\pm}=H_{21}^{\pm}=0$, as was the case in Section~\ref{sec:Nf1}. Hence, as far as complete $\cN$-fold SUSY systems satisfying both (\ref{eq:Nf1}) and (\ref{eq:Nf2}) are concerned, it is not necessary to solve them any more.

At this stage, it is interesting to compare the present results with the ones constructed by the universal enveloping algebra of $\fosp(2/2)$ in Section~\ref{sec:osp22}. We immediately see that, the $(2,2)$-component coefficients are completely identical with each other; $A(z)$ in (\ref{eq:A2}) coincides with the corresponding $A^{\fosp}(z)$ in (\ref{eq:Aosp}), $B_{22}^{-}(z)$ in (\ref{eq:B22-2}) with $B_{22}^{\fosp}(z)$ in (\ref{eq:Bosp11}), and $C_{22}^{-}(z)$ in (\ref{eq:C22-2}) with $C_{11}^{\fosp}(z)$ in (\ref{eq:Cosp11}), respectively, by the following parameter relations:
\begin{align}
\begin{split}
b_{2,22}+\frac{a_{3}}{2}=b_{+}-\frac{\cN-1}{2}b_{+J},\qquad b_{22,1}+a_{2}=b_{0}-\frac{\cN-1}{2}b_{0J},\\
b_{22,0}+\frac{a_{1}}{2}=b_{-}-\frac{\cN-1}{2}b_{-J},\quad c_{0}+\frac{\cN-1}{2}(b_{22,1}+a_{0})=b_{I}-\frac{\cN-1}{2}b_{J}.
\end{split}
\end{align}
On the contrary, however, the $(1,1)$-component coefficients are different from each other; $B_{11}^{-}(z)$ in (\ref{eq:B11-2}) and $C_{11}^{-}(z)$ in (\ref{eq:C11-2}), do not coincide with $B_{11}^{\fosp}(z)$ in (\ref{eq:Bosp11}) and $C_{11}^{\fosp}(z)$ in (\ref{eq:Cosp11}), respectively. 

The latter disagreement can be understood if we come back to $\tbP_{\cN}^{-}$ in the $\alpha(z)=0$ case to reconsider the solvable sector $\tbcV_{\cN}^{-}$ annihilated by it. In fact, for the diagonal $\tbP_{\cN}^{-}$ with $(1,1)$-component being a gauged type B $\cN$-fold supercharge component $\tP_{\cN}^{(\rmB)}$, it is not given by (\ref{eq:VN2}) but instead by
\begin{align}
\ker\tbP_{\cN}^{-}=\begin{pmatrix} \tcV_{\cN}^{(\rmB)}\\ \tcV_{\cN}^{(\rmA)}\end{pmatrix}\supset\tbcV_{\cN}^{-},
\label{eq:VN3}
\end{align}
where $\tcV_{\cN}^{(\rmB)}$ is an $\cN$-dimensional linear function space annihilated by $\tP_{\cN}^{(\rmB)}$ and, in particular for the above choice of $\beta(z)=-1/z$, is a type B monomial subspace given by
\begin{align}
\tcV_{\cN}^{(\rmB)}=\braket{1,z,\dots,z^{\cN-2},z^{\cN}}.
\end{align}
Therefore, $\ker\tbP_{\cN}^{-}\neq\tcV_{\cN}^{-}$ for (\ref{eq:VN2}) and (\ref{eq:tbP2}), which is exactly the cause of the disagreement. The finally obtained matrix $\cN$-fold SUSY is diagonal and is just a direct sum of one scalar type B and one scalar type A $\cN$-fold SUSY systems. Therefore, we shall hereafter assume that $\alpha(z)\neq 0$.

For arbitrary $\alpha(z)\neq 0$ and $\beta(z)$, we can easily show that
\begin{align}
\ker\tbP_{\cN}^{-}=\tbcV_{\cN}^{-}+\braket{\tbpsi(z)},
\label{eq:tkerP}
\end{align}
where the vector $\tbpsi(z)$ for $\beta(z)\neq 0$ is given by the solution to
\begin{align}
\tbpsi^{(\cN)}(z)=\rme^{-\int\rmd z\beta(z)}\begin{pmatrix} \beta(z)\\ \alpha(z) \end{pmatrix},
\label{eq:adVN1}
\end{align}
and for $\beta(z)=0$ by
\begin{align}
\tbpsi(z)= \begin{pmatrix} z^{\cN-1}\\ -(\cN-1)!\left(\int\rmd z\right)^{\cN}\alpha(z) \end{pmatrix}.
\label{eq:adVN2}
\end{align}
Hence, the operator $\tbP_{\cN}^{-}$ in (\ref{eq:tbP2}) is incomplete to identify the linear space $\tbcV_{\cN}^{-}$ in (\ref{eq:VN2}), although $\ker\tbP_{\cN}^{-}\supset\tbcV_{\cN}^{-}$. A crucial consequence is that the gauged matrix Hamiltonian $\tbH^{-}$ constructed by solving the gauged intertwining relation with respect to $\tbP_{\cN}^{-}$ preserves $\ker\tbP_{\cN}^{-}$ in (\ref{eq:tkerP}) but not necessarily $\tbcV_{\cN}^{-}$. It means in particular that the obtained $\tbH^{-}$ would contain a part which maps a subspace of $\tbcV_{\cN}^{-}$ outside $\tbcV_{\cN}^{-}$ though inside $\ker\tbP_{\cN}^{-}$. In practice, however, we can just remove \emph{a posteriori} the part by hand, as we shall demonstrate later, and no substantial trouble seems to arise.

The remaining concern is thus to solve the gauged intertwining relation directly. Due to the rich structure of $\tbP_{\cN}^{-}$, it turns to be much complicated, so we shall in this paper restrict ourselves to considering the particular case of $\beta(z)=0$ for the simplicity. We expect that the latter restriction would not so spoil the generality since the off-diagonal character of $\tbP_{\cN}$ is still intact, in contrast to the diagonal $\alpha(z)=0$ case studied above. Henceforth, we always assume $\beta(z)=0$. If we put the pair of gauged matrix Hamiltonians $\tbH^{\pm}$ as
\begin{align}
\tbH^{\pm}=\begin{pmatrix} -A\del_{z}^{2}-B_{11}^{\pm}\del_{z}-C_{11}^{\pm} & -B_{12}^{\pm}\del_{z}-C_{12}^{\pm}\\ -B_{21}^{\pm}\del_{z}-C_{21}^{\pm} & -A\del_{z}^{2}-B_{22}^{\pm}\del_{z}-C_{22}^{\pm}\end{pmatrix},
\label{eq:tHpmc}
\end{align}
the gauged version of the intertwining relation (\ref{eq:Nf1}) consists of four identities in its components given by
\begin{align}
\lefteqn{
\del_{z}^{\cN}\left( A\del_{z}^{2}+B_{11}^{-}\del_{z}+C_{11}^{-}\right)-\alpha\left[ B_{12}^{+}\left( \del_{z}+\frac{\cN A'}{2A}+\frac{\alpha'}{\alpha}\right)+C_{12}^{+}\right] \del_{z}^{\cN-1}
}\notag\\
&-\left[ A\left( \del_{z}+\frac{\cN A'}{2A}\right)^{2}+B_{11}^{+}\left( \del_{z}+\frac{\cN A'}{2A}\right) +C_{11}^{+}\right] \del_{z}^{\cN}=0,
\label{eq:int11}\\
&\del_{z}^{\cN}\left( B_{12}^{-}\del_{z}+C_{12}^{-}\right)-\left[ B_{12}^{+}\left( \del_{z}+\frac{\cN A'}{2A}\right) +C_{12}^{+}\right] \del_{z}^{\cN}=0,\\
\lefteqn{
\del_{z}^{\cN}\left(B_{21}^{-}\del_{z}+C_{21}^{-}\right)+\alpha\del_{z}^{\cN-1}\left( A\del_{z}^{2}+B_{11}^{-}\del_{z}+C_{11}^{-}\right)-\alpha\Biggl[ A\left(\del_{z}+\frac{\cN A'}{2A}+\frac{\alpha'}{\alpha}\right)^{2}
}\notag\\
&+B_{22}^{+}\left( \del_{z}+\frac{\cN A'}{2A}+\frac{\alpha'}{\alpha}\right)+C_{22}^{+}\Biggr] \del_{z}^{\cN-1}-\left[ B_{21}^{+}\left( \del_{z}+\frac{\cN A'}{2A}\right)+C_{21}^{+}\right] \del_{z}^{\cN}=0,\\
\lefteqn{
\del_{z}^{\cN}\left( A\del_{z}^{2}+B_{22}^{-}\del_{z}+C_{22}^{-}\right) +\alpha\del_{z}^{\cN-1}\left(B_{12}^{-}\del_{z}+C_{12}^{-}\right)
}\notag\\
&-\left[ A\left(\del_{z}+\frac{\cN A'}{2A}\right)^{2}+B_{22}^{+}\left(\del_{z}+\frac{\cN A'}{2A}\right)+C_{22}^{+}\right] \del_{z}^{\cN}=0.
\label{eq:int22}
\end{align}
If we put the products of $\cN$-fold supercharge components $\tbP_{\cN}^{\mp}\tbP_{\cN}^{\pm}$ appeared in the l.h.s.\ of the algebraic constraint (\ref{eq:Nf2}) as
\begin{align}
\tbP_{\cN}^{\mp}\tbP_{\cN}^{\pm}&=(-2A)^{\cN}\begin{pmatrix} \tP_{\cN,11}^{\pm} & \tP_{\cN,12}^{\pm}\\ \tP_{\cN,21}^{\pm} & \tP_{\cN,22}^{\pm} \end{pmatrix},
\label{eq:PNPN}
\end{align}
then each component $P_{\cN,ij}^{\pm}$ is calculated respectively as
\begin{align}
P_{\cN,11}^{+}&=(\del_{\cN, z})^{\cN}D_{z}^{\cN},
\label{eq:PNp11}\\
P_{\cN,12}^{+}&=-(\del_{\cN, z})^{\cN}D_{z}^{\cN-1}\alpha,\\
P_{\cN,21}^{+}&=\alpha(\del_{\cN, z})^{\cN-1}D_{z}^{\cN},\\
P_{\cN,22}^{+}&=( \del_{\cN, z})^{\cN}D_{z}^{\cN}-\alpha(\del_{\cN, z})^{\cN-1}D_{z}^{\cN-1}\alpha,
\end{align}
and
\begin{align}
P_{\cN,11}^{-}&=(D_{\cN, z})^{\cN}\del_{z}^{\cN}-(D_{\cN, z})^{\cN-1}\alpha^{2}\del_{z}^{\cN-1},\\
P_{\cN,12}^{-}&=-(D_{\cN, z})^{\cN-1}\alpha\del_{z}^{\cN},\\
P_{\cN,21}^{-}&=(D_{\cN, z})^{\cN}\alpha\del_{z}^{\cN-1},\\
P_{\cN,22}^{-}&=(D_{\cN, z})^{\cN}\del_{z}^{\cN},
\label{eq:PNm22}
\end{align}
where $\del_{\cN,z}$ and $D_{\cN,z}$ are given by (\ref{eq:dNz}). In spite of the simplification by setting $\beta(z)=0$, the set of the intertwining relations (\ref{eq:int11})--(\ref{eq:int22}) is still too complicated to obtain general solutions for an arbitrary $\cN\in\bbN$. In the followings, we thus content ourselves with examining a few lower $\cN$ cases.

\subsection{The $\cN=1$ Case}
\label{ssec:N=1}

In the case of $\cN=1$, namely, of ordinary SUSY, the set of intertwining relations (\ref{eq:int11})--(\ref{eq:int22}) can be integrated completely, and we have for $B_{ij}^{\pm}(z)$
\begin{align}
B_{ij}^{-}=B_{ij}^{+}:=B_{ij}\quad (i, j=1, 2),
\label{eq:B1pm}
\end{align}
for $C_{ij}^{+}(z)$
\begin{align}
C_{11}^{+}=&-\frac{A''}{2}+\frac{(A')^{2}}{4A}-\frac{B_{11}A'}{2A}+B'_{11}+C_{12}^{(0)}\int\rmd z\,\alpha+C_{11}^{(0)},
\label{eq:C1p11}\\
C_{12}^{+}=&-\frac{B_{12}A'}{2A}+B'_{12}+C_{12}^{(0)},\\
C_{21}^{+}=&-\frac{B_{21}A'}{2A}+B'_{21}-A'\alpha-A\alpha'+B_{11}\alpha-2C_{12}^{(0)}\int\rmd z\left( \alpha\int\rmd z\,\alpha\right)\notag\\
&+(C_{22}^{(0)}-C_{11}^{(0)})\int\rmd z\,\alpha+C_{21}^{(0)},\\
C_{22}^{+}=&-\frac{A''}{2}+\frac{(A')^{2}}{4A}-\frac{B_{22}A'}{2A}+B'_{22}+B_{12}\alpha-C_{12}^{(0)}\int\rmd z\,\alpha+C_{22}^{(0)},
\end{align}
and for $C_{ij}^{-}(z)$
\begin{align}
C_{11}^{-}=&\;B_{12}\alpha+C_{12}^{(0)}\int\rmd z\,\alpha+C_{11}^{(0)},
\label{eq:C1m11}\\
C_{12}^{-}=&\;C_{12}^{(0)},\\
C_{21}^{-}=&\;A\alpha'+B_{22}\alpha-2C_{12}^{(0)}\int\rmd z\left( \alpha\int\rmd z\,\alpha\right)\notag\\
&+(C_{22}^{(0)}-C_{11}^{(0)})\int\rmd z\,\alpha+C_{21}^{(0)},
\label{eq:C1m21}\\
C_{22}^{-}=&-C_{12}^{(0)}\int\rmd z\,\alpha+C_{22}^{(0)},
\label{eq:C1m22}
\end{align}
where $C_{ij}^{(0)}$ are integral constants. In this $\cN=1$ case, we note that the additional element $\tbpsi(z)$ of $\ker\tbP_{\cN}^{-}$ given in (\ref{eq:adVN2}) reads as
\begin{align}
\tbpsi(z) =\begin{pmatrix} 1\\ -\int\rmd z\,\alpha(z) \end{pmatrix}.
\end{align}
Then, we can easily check that
\begin{align}
\tbH^{-}\begin{pmatrix} 0\\ 1\end{pmatrix}=-C_{12}^{(0)}\tbpsi(z) -C_{22}^{(0)}\begin{pmatrix} 0\\ 1\end{pmatrix},
\label{eq:Hop1}
\end{align}
and that the part of $\tbH^{-}$ which maps an element of $\tbcV_{1}^{-}$ outside it but inside $\ker\tbP_{1}^{-}$ comes from the term proportional to $C_{12}^{(0)}$. Hence, $\tbH^{-}$ preserves the original space $\tbcV_{1}^{-}$ only if we put $C_{12}^{(0)}=0$. The resulting system contains 6 arbitrary functions $A(z)$, $B_{ij}(z)$ ($i,j=1,2$), and $\alpha(z)$, and 3 free parameters $C_{11}^{(0)}$, $C_{21}^{(0)}$, and $C_{22}^{(0)}$. Comparing $\tbH^{-}$ here with the $\fosp(2/2)$ model $\tbH^{\fosp}$ for $\cN=1$, whose constraints are given by (\ref{eq:ospN1}), we observe that in contrast with the arbitrariness of $C_{11}^{\fosp}(z)$ and $C_{21}^{\fosp}(z)$ in the latter, $C_{11}^{-}(z)$ and $C_{21}^{-}$ in the former have particular functional forms shown in (\ref{eq:C1m11}) and (\ref{eq:C1m21}), although all the other coefficients are identical in the both with $C_{22}^{(0)}=c_{0}$. The difference is due to the fact that the former maps $\tbpsi(z)$ into $\ker\tbP_{1}^{-}$. Indeed, we can check that
\begin{align}
\tbH^{-}\tbpsi(z)=-C_{11}^{(0)}\tbpsi(z)-C_{21}^{(0)}\begin{pmatrix} 0\\ 1\end{pmatrix}.
\end{align}
Hence, $\tbH^{-}$ with the constraint $C_{12}^{(0)}$ preserves not only $\tbcV_{1}^{-}$ but also a finite flag of the two spaces $\tbcV_{1}^{-}\subset\ker\tbP_{1}^{-}$. That is why $\tbH^{-}$ falls into a particular case of $\tbH^{\fosp}$ for $\cN=1$.

Next, we shall impose the transposition symmetry. It follows from the first condition (\ref{eq:tB4}) that
\begin{align}
B_{11}=B_{22}:=B,\qquad B_{21}=-B_{12}.
\label{eq:tpc1}
\end{align}
Applying the second condition (\ref{eq:tC2}) to the above formulas for $C_{ij}^{\pm}(z)$, we obtain
\begin{align}
B_{12}=-A\alpha,
\label{eq:tpc2}\\
2C_{12}^{(0)}\int\rmd z\left( \alpha\int\rmd z\,\alpha\right) +(C_{11}^{(0)}-C_{22}^{(0)})\int\rmd z\,\alpha+C_{12}^{(0)}-C_{21}^{(0)}=0.
\label{eq:tpc3}
\end{align}
Differentiating (\ref{eq:tpc3}) with respect to $z$ yields $2C_{12}^{(0)}\int\rmd z\,\alpha=C_{22}^{(0)}-C_{11}^{(0)}$, which means under the prerequisite $\alpha(z)\neq 0$ that $C_{12}^{(0)}=C_{11}^{(0)}-C_{22}^{(0)}=0$. Substituting them back into (\ref{eq:tpc3}), we finally obtain
\begin{align}
C_{12}^{(0)}=C_{21}^{(0)}=0,\qquad C_{11}^{(0)}=C_{22}^{(0)}.
\label{eq:tpc4}
\end{align}

At last, let us examine the algebraic constraint (\ref{eq:Nf2}). From the formulas (\ref{eq:PNp11})--(\ref{eq:PNm22}) for $\cN=1$, the matrix elements of the product operators (\ref{eq:PNPN}) read as
\begin{align}
\tP_{1,11}^{+}&=\del_{z}^{2}+\frac{2Q+A'}{2A}\del_{z}+\frac{Q'}{A}-\frac{QA'}{2A^{2}},
\label{eq:tP1p11}\\
\tP_{1,12}^{+}&=-\alpha\del_{z}-\frac{A'}{2A}\alpha-\alpha',\\
\tP_{1,21}^{+}&=\alpha\del_{z}+\frac{Q}{A}\alpha,\\
\tP_{1,22}^{+}&=\del_{z}^{2}+\frac{2Q+A'}{2A}\del_{z}+\frac{Q'}{A}-\frac{QA'}{2A^{2}}-\alpha^{2},
\end{align}
and
\begin{align}
\tP_{1,11}^{-}&=\del_{z}^{2}+\frac{2Q+A'}{2A}\del_{z}-\alpha^{2},\\
\tP_{1,12}^{-}&=-\alpha\del_{z},\\
\tP_{1,21}^{-}&=\alpha\del_{z}+\alpha'+\frac{2Q+A'}{2A}\alpha,\\
\tP_{1,22}^{-}&=\del_{z}^{2}+\frac{2Q+A'}{2A}\del_{z}.
\label{eq:tP1m22}
\end{align}
Comparing the coefficients (\ref{eq:B1pm})--(\ref{eq:C1m22}) of the component Hamiltonians $H_{ij}^{\pm}$ defined via (\ref{eq:tHpmc}) with the ones in (\ref{eq:tP1p11})--(\ref{eq:tP1m22}), and noting the transposition symmetry of the constant matrix $\bC_{0}$ (cf.\ (\ref{eq:Ccon})), we find that the algebraic relation (\ref{eq:Nf2}) holds, if and only if $\bC_{0}=C_{0}\bI_{2}$ and, for $B_{ij}(z)$
\begin{align}
B_{11}=B_{22}=\frac{2Q+A'}{2},\qquad B_{12}=-B_{21}=-A\alpha,
\label{eq:alg1}
\end{align}
and for $C_{ij}^{\pm}(z)$
\begin{align}
C_{12}^{(0)}=C_{21}^{(0)}=0,\qquad C_{11}^{(0)}=C_{22}^{(0)}=-C_{0}.
\label{eq:alg2}
\end{align}
{}From the relation (\ref{eq:defQ}) for $\cN=1$, we see that the constraints (\ref{eq:alg1}) and  (\ref{eq:alg2}) are completely identical to the conditions for transposition symmetry (\ref{eq:tpc1}), (\ref{eq:tpc2}), and (\ref{eq:tpc4}). It shows that the intertwining relation (\ref{eq:Nf1}) with transposition symmetry is equivalent to the algebraic relation (\ref{eq:Nf2}) in this case.

\subsection{The $\cN=2$ Case}
\label{ssec:N=2}

In the case of 2-fold SUSY, we can also solve the set of intertwining relations (\ref{eq:int11})--(\ref{eq:int22}) analytically. As in the $\cN=1$ case, the plus and minus coefficients of first order $B_{ij}^{\pm}(z)$ are identical with each other in all the components:
\begin{align}
B_{ij}^{-}=B_{ij}^{+}:=B_{ij}\quad (i, j=1, 2).
\label{eq:B2pm}
\end{align}
The plus coefficients of zeroth order $C_{ij}^{+}(z)$ are given by
\begin{align}
C_{11}^{+}&=-\frac{B_{11}A'}{A}+2B'_{11}-B_{12}\alpha+C_{11}^{-},
\label{eq:C2p11}\\
C_{12}^{+}&=-\frac{B_{12}A'}{A}+2B'_{12}+C_{12}^{-},\\
C_{21}^{+}&=-A'\alpha-2A\alpha'-\frac{B_{21}A'}{A}+2B'_{21}-(B_{22}-B_{11})\alpha+C_{21}^{-},\\
C_{22}^{+}&=-\frac{B_{22}A'}{A}+2B'_{22}+B_{12}\alpha+C_{22}^{-}.
\end{align}
The minus coefficients $C_{ij}^{-}(z)$ are integrated as
\begin{align}
C_{1i}^{-}=C_{1i}^{(1)}z+C_{1i}^{(0)},\quad 
C_{2i}^{-}=-C_{1i}^{(1)}\int\rmd z\int\rmd z\,\alpha+C_{2i}^{(1)}z+C_{2i}^{(0)}\quad (i=1,2),
\label{eq:C2m1}
\end{align}
where $C_{ij}^{(1)}$ and $C_{ij}^{(0)}$ are integral constants. In contrast to the $\cN=1$ case, the first-order coefficients cannot be arbitrary but must satisfy
\begin{align}
\begin{split}
B_{11}=&-C_{12}^{(1)}z\int\rmd z\,\alpha z+C_{12}^{(0)}\int\rmd z\int\rmd z\,\alpha +(B_{12}^{(1)}z+B_{12}^{(0)})\int\rmd z\,\alpha\\
&-C_{11}^{(1)}z^{2}+B_{11}^{(1)}z+B_{11}^{(0)},\\
B_{12}=&-C_{12}^{(1)}z^{2}+B_{12}^{(1)}z+B_{12}^{(0)},\\
B_{21}=&\left[ \int\rmd z\,\alpha (C_{12}^{(1)}z-B_{12}^{(1)}-C_{12}^{(0)})+C_{11}^{(1)}z-B_{11}^{(1)}-C_{11}^{(0)}+C_{22}^{(0)}\right] \int\rmd z\int\rmd z\,\alpha\\
&+A\alpha -C_{22}^{(1)}z\int\rmd z\,\alpha z+(B_{22}^{(1)}z+B_{22}^{(0)})\int\rmd z\,\alpha-C_{21}^{(1)}z^{2}+B_{21}^{(1)}z+B_{21}^{(0)},\\
B_{22}=&(C_{12}^{(1)}z-B_{12}^{(1)}-C_{12}^{(0)})\int\rmd z\int\rmd z\,\alpha-C_{22}^{(1)}z^{2}+B_{22}^{(1)}z+B_{22}^{(0)},
\end{split}
\label{eq:B2''}
\end{align}
where $B_{ij}^{(1)}$ and $B_{ij}^{(0)}$ ($i,j=1,2$) are integral constants. In the $\cN=2$ case, the additional element $\tbpsi(z)$ of $\ker\tbP_{\cN}^{-}$ given in (\ref{eq:adVN2}) reads as
\begin{align}
\tbpsi(z) =\begin{pmatrix} z\\ -\int\rmd z\int\rmd z\,\alpha(z) \end{pmatrix}.
\end{align}
Operating $\tbH^{-}$ on the basis of $\tbcV_{2}$, we have
\begin{align}
\begin{split}
\tbH^{-}\begin{pmatrix} 1\\ 0\end{pmatrix}&=-C_{11}^{(1)}\tbpsi(z) -C_{11}^{(0)}\begin{pmatrix} 1\\ 0\end{pmatrix}-C_{21}^{(1)}\begin{pmatrix} 0\\ z\end{pmatrix}-C_{21}^{(0)}\begin{pmatrix} 0\\ 1\end{pmatrix},\\
\tbH^{-}\begin{pmatrix} 0\\ z\end{pmatrix}&=-(B_{12}^{(1)}+C_{12}^{(0)})\tbpsi(z) -B_{12}^{(0)}\begin{pmatrix} 1\\ 0\end{pmatrix}-(B_{22}^{(1)}+C_{22}^{(0)})\begin{pmatrix} 0\\ z\end{pmatrix}-B_{22}^{(0)}\begin{pmatrix} 0\\ 1\end{pmatrix},\\
\tbH^{-}\begin{pmatrix} 0\\ 1\end{pmatrix}&=-C_{12}^{(1)}\tbpsi(z) -C_{12}^{(0)}\begin{pmatrix} 1\\ 0\end{pmatrix}-C_{22}^{(1)}\begin{pmatrix} 0\\ z\end{pmatrix}-C_{22}^{(0)}\begin{pmatrix} 0\\ 1\end{pmatrix},
\end{split}
\end{align}
and find that the part of $\tbH^{-}$ which maps an element of $\tbcV_{2}^{-}$ outside it but inside $\ker\tbP_{2}^{-}$ comes from the terms proportional to $C_{11}^{(1)}$, $B_{12}^{(1)}+C_{12}^{(0)}$, and $C_{12}^{(0)}$. Hence, $\tbH^{-}$ preserves the original space $\tbcV_{1}^{-}$ only if 
\begin{align}
C_{11}^{(1)}=B_{12}^{(1)}+C_{12}^{(0)}=C_{12}^{(1)}=0.
\label{eq:const2}
\end{align}
Substituting them into (\ref{eq:C2m1}) and solving (\ref{eq:B2''}), we obtain the final form of the coefficients as
\begin{align}
C_{1i}^{-}=&\;C_{1i}^{(0)},\quad C_{2i}^{-}=C_{2i}^{(1)}z+C_{2i}^{(0)}\quad (i=1,2),
\label{eq:C2m2}\\
B_{11}=&\;-C_{12}^{(0)}\int\rmd z\,\alpha z+B_{12}^{(0)}\int\rmd z\,\alpha+B_{11}^{(1)}z+B_{11}^{(0)},
\label{eq:B2m11}\\
B_{12}=&\;-C_{12}^{(0)}z+B_{12}^{(0)},\\
B_{21}=&-(B_{11}^{(1)}+C_{11}^{(0)}-C_{22}^{(0)})\int\rmd z\int\rmd z\,\alpha+A\alpha-C_{22}^{(1)}z\int\rmd z\,\alpha z\notag\\
&+(B_{22}^{(1)}z+B_{22}^{(0)})\int\rmd z\,\alpha -C_{21}^{(1)}z^{2}+B_{21}^{(1)}z+B_{21}^{(0)},
\label{eq:B2m21}\\
B_{22}=&-C_{22}^{(1)}z^{2}+B_{22}^{(1)}z+B_{22}^{(0)}.
\label{eq:B2m22}
\end{align}
Hence, the system contains 2 arbitrary functions $A(z)$ and $\alpha(z)$, and 13 free parameters $C_{1i}^{(0)}$, $C_{2i}^{(1)}$, $C_{2i}^{(0)}$, $B_{ii}^{(1)}$, $B_{ii}^{(0)}$ ($i=1,2$), $B_{21}^{(1)}$, $B_{21}^{(0)}$, and $B_{12}^{(0)}$. Comparing $\tbH^{-}$ here with the $\fosp(2/2)$ model $\tbH^{\fosp}$ for $\cN=2$, whose constraints are given by (\ref{eq:ospN2}), we see that in contrast with the arbitrariness of $B_{11}^{\fosp}(z)$ and $B_{21}^{\fosp}(z)$ in the latter, $B_{11}^{-}(z)$ and $B_{21}^{-}$ in the former have particular functional forms shown in (\ref{eq:B2m11}) and (\ref{eq:B2m21}). All the other coefficients are identical in the both with the following parameter relations:
\begin{align}
\begin{split}
C_{11}^{(0)}&=b_{I}-b_{J},\quad C_{12}^{(0)}=-\bar{f}_{+},\quad C_{21}^{(1)}=f_{+},\quad C_{21}^{(0)}=f_{-},\quad C_{22}^{(1)}=-\frac{b_{+}}{2},\\
C_{22}^{(0)}&=-\frac{b_{0}}{4}-\frac{b_{J}}{2}+b_{I},\quad B_{12}^{(0)}=-\bar{f}_{-},\quad B_{22}^{(1)}=\frac{b_{0}}{2},\quad B_{22}^{(0)}=\frac{b_{-}}{2}.
\end{split}
\end{align}
As in the case of $\cN=1$, the difference comes from the fact that the former maps $\tbpsi(z)$ into $\ker\tbP_{2}^{-}$. Actually, operating $\tbH^{-}$ on $\tbpsi(z)$, we have
\begin{align}
\tbH^{-}\tbpsi(z)=-(B_{11}^{(1)}+C_{11}^{(0)})\tbpsi(z)-B_{11}^{(0)}\begin{pmatrix} 1\\ 0\end{pmatrix}-(B_{21}^{(1)}+C_{21}^{(0)})\begin{pmatrix} 0\\ z\end{pmatrix}-B_{21}^{(0)}\begin{pmatrix} 0\\ 1\end{pmatrix}.
\end{align}
Hence, $\tbH^{-}$ with the constraint (\ref{eq:const2}) preserves not only $\tbcV_{2}^{-}$ but also a finite flag of the two spaces $\tbcV_{2}^{-}\subset\ker\tbP_{2}^{-}$. That is again the reason why $\tbH^{-}$ is realized as a particular case of $\tbH^{\fosp}$ for $\cN=2$.

Next, we shall impose the transposition symmetry. It is obvious that the first condition (\ref{eq:tB4}) leads to the same one as (\ref{eq:tpc1}):
\begin{align}
B_{11}=B_{22}:=B,\qquad B_{21}=-B_{12}.
\label{eq:tpc5}
\end{align}
Applying the second condition (\ref{eq:tC2}) to the above formulas for $C_{ij}^{\pm}(z)$, we obtain
\begin{align}
2B_{12}&=-A\alpha,
\label{eq:tpc6}\\
2C_{21}^{-}&=2C_{12}^{-}+A\alpha'+B\alpha.
\label{eq:tpc7}
\end{align}
One can easily see that the conditions (\ref{eq:tpc5})--(\ref{eq:tpc7}) impose quite strong constraints on the form of the coefficients when (\ref{eq:C2m2})--(\ref{eq:B2m22}) are substituted into them. As we shall show shortly, they are consistent with the algebraic relation (\ref{eq:Nf2}) where the transposition symmetry is already incorporated.

At last, let us examine the algebraic constraint (\ref{eq:Nf2}). The matrix elements of the product operators (\ref{eq:PNPN}) can be calculated by using the formulas (\ref{eq:PNp11})--(\ref{eq:PNm22}) for $\cN=2$ but we here omit their expressions since they are lengthy and complex even after the simplification of $\beta(z)=0$. Both sides of (\ref{eq:Nf2}) are of fourth order in the $\cN=2$ case, and each coefficient of differential operators of the same order must be identical in the both sides. The fourth-order terms are already equal, and equating the third-order terms yields
\begin{align}
B_{11}=B_{22}=Q,\qquad 2B_{12}=-2B_{21}=-A\alpha.
\label{eq:alg3}
\end{align}
We note here that the constraint (\ref{eq:alg3}) is totally identical to the conditions (\ref{eq:tpc5}) and (\ref{eq:tpc6}) for the transposition symmetry via the relation $Q(z)=B(z)$ given by (\ref{eq:defQ}) for $\cN=2$. Using these identities and equating the coefficients of second order, we obtain the identities for $C_{ij}^{+}(z)$
\begin{align}
\begin{split}
C_{11}^{+}&=\frac{3}{2}Q'-\frac{A'Q}{A}+\frac{1}{8}A\alpha^{2}+C_{0,11},\quad 
C_{12}^{+}=-\frac{1}{4}A'\alpha-A\alpha'+C_{0,12},\\
C_{21}^{+}&=\frac{1}{2}Q\alpha-\frac{1}{4}A'\alpha-\frac{1}{2}A\alpha'+C_{0,21},\quad 
C_{22}^{+}=\frac{3}{2}Q'-\frac{A'Q}{A}-\frac{3}{8}A\alpha^{2}+C_{0,22},
\end{split}
\label{eq:C2p}
\end{align}
and for $C_{ij}^{-}(z)$
\begin{align}
\begin{split}
C_{11}^{-}&=-\frac{1}{2}Q'-\frac{3}{8}A\alpha^{2}+C_{0,11},\quad 
C_{12}^{-}=\frac{1}{4}A'\alpha+C_{0,12},\\
C_{21}^{-}&=\frac{1}{2}Q\alpha+\frac{1}{4}A'\alpha+\frac{1}{2}A\alpha'+C_{0,21},\quad 
C_{22}^{-}=-\frac{1}{2}Q'+\frac{1}{8}A\alpha^{2}+C_{0,22}.
\end{split}
\label{eq:C2m}
\end{align}
We note that the above formulas automatically satisfy the last condition (\ref{eq:tpc7}) for the transposition symmetry of $\tbH^{-}$ by taking into account $C_{0,21}=C_{0,12}$, the assumed transposition symmetry of $\bC_{0}$ in (\ref{eq:Ccon}). Hence, all the conditions (\ref{eq:tpc5})--(\ref{eq:tpc7}) already hold by equating the coefficients of the third and second orders in the algebraic relation (\ref{eq:Nf2}) for $\cN=2$.

On the other hand, equating the first- and the zeroth-order terms results in constraints among $A(z)$, $Q(z)$, and $\alpha(z)$. From the first order, we have
\begin{align}
A'\alpha'+A\alpha''=0,\qquad 4Q\alpha'-A\alpha^{3}=0,
\label{eq:aqab1}
\end{align}
while from the zeroth order
\begin{align}
4Q^{(3)}-3A''\alpha^{2}-6A'\alpha\alpha'-2A\left[ \alpha\alpha''+(\alpha')^{2}\right]=0,
\label{eq:aqab2}\\
A^{(3)}\alpha+2A''\alpha'+A'\alpha''=0,
\label{eq:aqab4}\\
4Q\alpha''+A'\alpha^{3}=0,
\label{eq:aqab5}\\
3A''\alpha^{2}+6A'\alpha\alpha'+A\left[ \alpha\alpha''+4(\alpha')^{2}\right]=0.
\label{eq:aqab6}
\end{align}
{}From these constraints, we can reproduce the formulas (\ref{eq:C2m1}) which were obtained by solving the intertwining relation. First, integrating twice (\ref{eq:aqab4}), we have
\begin{align}
A'\alpha=C_{1}z+C_{2}.
\label{eq:integ1}
\end{align}
Using this and (\ref{eq:aqab2}) yields $4Q^{(3)}-(A\alpha^{2})'=2C_{1}\alpha$ and thus
\begin{align}
4Q'-A\alpha^{2}=2C_{1}\int\rmd z\int\rmd z\,\alpha+C_{3}z+C_{4}.
\end{align}
Next, combining (\ref{eq:aqab1}), (\ref{eq:aqab2}) and (\ref{eq:aqab6}), we obtain $4Q^{(3)}+3(A\alpha^{2})''=0$ and thus
\begin{align}
4Q'+3A\alpha^{2}=C_{5}z+C_{6}.
\end{align}
Then, using the latter, (\ref{eq:aqab1}), and (\ref{eq:aqab5}), we can show that
\begin{align}
4Q\alpha+4A\alpha'=C_{5}\int\rmd z\int\rmd z\,\alpha+C_{7}z+C_{8}.
\label{eq:integ2}
\end{align}
In the above, $C_{i}$ ($i=1,\dots,8$) are integral constants. Substituting (\ref{eq:integ1})--(\ref{eq:integ2}) into (\ref{eq:C2p}) and (\ref{eq:C2m}), and comparing the resultant formulas with (\ref{eq:C2p11})-(\ref{eq:C2m1}), we see that they are entirely identical with each other, connected by the following parameter relations:
\begin{align}
\begin{split}
8C_{11}^{(1)}&=-C_{5},\qquad 8C_{11}^{(0)}=-C_{6}+8C_{0,11},\\
4C_{12}^{(1)}&=C_{1},\qquad 4C_{12}^{(0)}=C_{2}+4C_{0,12},\\
8C_{21}^{(1)}&=2C_{1}+C_{7},\qquad 8C_{21}^{(0)}=2C_{2}+C_{8}+8C_{0,21},\\
8C_{22}^{(1)}&=-C_{3},\qquad 8C_{22}^{(0)}=-C_{4}+8C_{0,22}.
\end{split}
\end{align}
Therefore, the system obtained from the algebraic constraint is essentially identical to the one from the intertwining relation for $\cN=2$..

The zeroth-order of the algebraic constraint (\ref{eq:Nf2}) in addition provides the expressions for the matrix elements of $\bC_{1}$ in terms of $A(z)$, $Q(z)$, and $\alpha(z)$ as
\begin{align}
8C_{1,11}=&\;8C_{1,22}+D_{2}[A,Q,\alpha],
\label{eq:C111}\\
16C_{1,12}=&\;16C_{1,21}=D_{1}[A,Q,\alpha] \alpha,
\label{eq:C112}\\
64C_{1,22}=&\;32QQ''-16(Q')^{2}+4\left[ 2AA''-(A')^{2}\right]\alpha^{2}\notag\\
&-8(2A'Q-AQ')\alpha^{2}-16AQ\alpha\alpha'-A^{2}\alpha^{4},
\label{eq:C122}
\end{align}
where $D_{1}$ and $D_{2}$ are defined by
\begin{align}
D_{1}[A,Q,\alpha ]&=-4(A''Q-A'Q'+AQ'')+AA'\alpha^{2}+2A^{2}\alpha\alpha',\\
D_{2}[A,Q,\alpha ]&=2(2A'Q-AQ')\alpha^{2}+2AQ\alpha\alpha' +A^{2}\alpha^{4}.
\end{align}
At first sight, it is not evident that each of the r.h.s.\ of (\ref{eq:C111})--(\ref{eq:C122}) is a constant. To show that it is actually the case, we first note that by multiplying (\ref{eq:aqab4}) by $-4Q$ and using (\ref{eq:aqab1}) and (\ref{eq:aqab5}) we obtain
\begin{align}
\left\{ -4A^{(3)}Q-\left[ 2AA''-(A')^{2}\right] \alpha^{2}\right\} \alpha=\frac{\rmd D_{1}}{\rmd z}\alpha=0.
\end{align}
Since $\alpha\neq 0$, we have $\rmd D_{1}/\rmd z=0$ and thus $D_{1}=C_{9}$(=constant). Multiplying the latter by $\alpha'$ and integrating the result with respect to $z$, on one hand, we obtain $D_{1}\alpha=C_{9}\alpha+C_{10}$, while multiplying it by $-2\alpha^{2}$ and integrating the result with respect to $z$, on the other hand, we obtain
\begin{align}
32QQ''-16(Q')^{2}+4\left[ 2AA''-(A')^{2}\right]\alpha^{2}-8(2A'Q-AQ')\alpha^{2}\notag\\
-16AQ\alpha\alpha'-A^{2}\alpha^{4}=-2C_{9}\int\rmd z\,\alpha^{2}+C_{11},
\end{align}
where $C_{10}$ and $C_{11}$ are integral constants. They are compatible with (\ref{eq:C112}) and (\ref{eq:C122}) only when $C_{9}=0$. Hence, we conclude that
\begin{align}
D_{1}[A,Q,\alpha ]=0.
\label{eq:D1=0}
\end{align}
Then, it is straightforward to show from (\ref{eq:aqab1})--(\ref{eq:aqab6}) and (\ref{eq:D1=0}) that
\begin{align}
\frac{\rmd}{\rmd z}D_{2}[A,Q,\alpha ]=0.
\end{align}
Hence, all the r.h.s.\ of (\ref{eq:C111})--(\ref{eq:C122}) are in fact constants, and in particular
$C_{1,12}=C_{1,21}=0$, that is, $\bC_{1}$ is diagonal. The second assumption for $\bC_{1}$ in (\ref{eq:Ccon}) now reads as $(C_{1,11}-C_{1,22})H_{12}^{\pm}=0$ which can be satisfied only if
\begin{align}
C_{1,11}=C_{1,22},\qquad \bC_{1}=C_{1,22}\bI_{2},\qquad D_{2}[A,Q,\alpha ]=0,
\end{align}
and $C_{1,22}$ is given by (\ref{eq:C122}).

\section{Discussion and Summary}
\label{sec:discus}

In this paper, we have developed the framework of $\cN$-fold SUSY in quantum mechanical matrix systems and applied it to construct models associated with the $\fq(2)$ and $\fosp(2/2)$ Lie-superalgebraic quasi-solvable models. In the $\fq(2)$ case, it turns out that the intertwining relation solely leads to the system which is identical to the $\fq(2)$ model. The closure of $\cN$-fold superalgebra is virtually equivalent to the combination of the intertwining and the transposition symmetry although the difference has been found for the non-trivial $Q_{A}(z)$. The former algebraic requirement imposes the strong constraint on the admissible form so that all the matrix operators in the system are proportional to the unit matrix and the resultant model is equivalent to a scalar type A $\cN$-fold SUSY one. In the $\fosp(2/2)$ case, the $\cN$-fold supercharge $\tbP_{\cN}^{-}$ admits two additional functions $\alpha(z)$ and $\beta(z)$ in the components due to the difference in the dimension of the upper and lower components of the originally considered invariant subspace $\tbcV_{\cN}^{-}$. The latter fact further admits another vector $\tbpsi(z)$ annihilated by the $\cN$-fold supercharge and thus extends the original invariant subspace by one dimension. In the particular case of $\alpha(z)=0$, the extended invariant subspace consists of an $\cN$-dimensional type B monomial subspace in the upper component and and $\cN$-dimensional type A monomial subspace in the lower component, and the resulting system as a consequence composed of one scalar type B and one scalar type A $\cN$-fold SUSY models. For the case of $\alpha(z)\neq 0$, we have restricted our investigation to the $\beta(z)=0$ for $\cN=1, 2$ because of the complexity of the intertwining relation. In both the $\cN=1$ and $\cN=2$ cases, the intertwining relation leads to particular cases of the $\fosp(2/2)$ quasi-solvable models due to the fact that the component Hamiltonian $\tbH^{-}$ preserves the additional vector $\tbpsi(z)$ in addition to $\tbcV_{\cN}^{-}$. The requirements of the transposition symmetry and of the closure of $\cN$-fold superalgebra result in the same constraints on the form of the system.

It is evident from the whole results, the framework of $\cN$-fold SUSY proves to be quite efficient in constructing quasi-solvable quantum mechanical matrix models, as in the case of ordinary scalar ones. It is also true, however, that various novel features which do not exist in the latter case emerge in the present analysis. First of all, the equivalence between weak quasi-solvability and $\cN$-fold SUSY which holds in the scalar case is apparently violated in the matrix case, which was already indicated in~\cite{Ta12a}. As we have already recognized throughout the examinations in this paper, the algebraic constraint (\ref{eq:Nf2}) is always stronger than the intertwining relation (\ref{eq:Nf1}) in all the cases. In this respect, it is worth noting the fact that the intertwining together with the transposition symmetry leads to almost the same condition as the algebraic constraint, at least in the present cases. For ordinary scalar Schr\"{o}dinger operators, the transposition symmetry is a built-in structure in a linear function space and plays a crucial role in closing $\cN$-fold superalgebra~\cite{AS03}. For $n\times n$ matrix Schr\"{o}dinger operators, it is not the case since they also act on an $n$-dimensional vector space where they have \emph{a priori} no particular symmetry. On the other hand, the whole $\cN$-fold SUSY structures (\ref{eq:Nf1}) and (\ref{eq:Nf2}) are interrelated to some extent so that each element $\tbH^{\pm}$ and $\tbP_{\cN}^{\pm}$ must transform consistently under certain symmetry transformations. That is exactly the reason why the notion of conjugation has been incorporated in this paper, which was obscure in~\cite{Ta12a}. Although we have only employed the transposition symmetry as a conjugation, it is interesting to examine other candidates such as the Hermitian conjugate (\ref{eq:Herm}) and the inverse (\ref{eq:inver}). It would be a challenging problem to prove whether the intertwining together with a certain conjugation is exactly equivalent to the algebraic constraint.

Another novel and intriguing aspect arises when there are differences in dimension among linear spaces in a multi-component invariant subspace $\tbcV_{\cN}^{-}$ under consideration. The closure of $\cN$-fold superalgebra for a $n\times n$ matrix Schr\"{o}dinger operator inevitably requires that every diagonal elements of $\cN$-fold supercharge $\tbP_{\cN}^{-}$ are of $\cN$th order while the other off-diagonal ones are of less than $\cN$th order. Then, all the elements of $\ker\tbP_{\cN}^{-}$ are determined by $n$ sets of $\cN$th-order linear differential equations.
Hence, the dimension of $\ker\tbP_{\cN}^{-}$ must be always equal to $n\cN$. In addition, any subspace of $\ker\tbP_{\cN}^{-}$ whose elements have only one non-zero component in the $n$-dimensional vector space is at most $\cN$ dimensional. The latter means that if we construct an $n$-component linear space $\tbcV_{\cN}^{-}$ to be preserved by $\tbH^{-}$ as a direct sum of one-component scalar linear spaces, $\tbcV_{\cN}^{-}=\tcV_{\cN_{1}}^{-}\be_{1}+\dots+\tcV_{\cN_{n}}^{-}\be_{n}$ where $\be_{k}$ is the $k$th unit vector and each dimension satisfies $0\leq\cN_{k}\leq\cN$, we always have $m=n\cN-\sum_{k=1}^{n}\cN_{k}$ additional vectors $\tbpsi_{i}$ ($i=1,\dots,m$) such that $\tbpsi_{i}\in\ker\tbP_{\cN}^{-}$ but $\tbpsi_{i}\not\in\tbcV_{\cN}^{-}$. In the $\fosp(2/2)$ case where $\tbcV_{\cN}^{-}$ is given by (\ref{eq:VN2}), $m=1$ and thus either the upper space $\tcV_{\cN-1}^{(\rmA)}$ is extended to $\tcV_{\cN}^{(\rmB)}$ for $\alpha(z)=0$ to yield (\ref{eq:VN3}) or the vector $\tbpsi(z)$ satisfying (\ref{eq:adVN1}) for $\beta(z)\neq 0$ or (\ref{eq:adVN2}) for $\beta(z)=0$ to yield (\ref{eq:tkerP}). Hence, it is not always possible to construct any Lie-superalgebraic quasi-solvable quantum mechanical matrix models, whose invariant subspace $\tbcV_{\cN}^{-}$ is given by a finite-dimensional module, by using the $\cN$-fold SUSY formulation. As demonstrated in Section~\ref{sec:Nf2}, we can remove unfavorable terms which does not keep $\tbcV_{\cN}^{-}$ invariant by hand such that $\tbH^{-}$ preserves a finite flag of the vector spaces $\tbcV_{\cN}^{-}\subset\ker\tbP_{\cN}^{-}$. The obtained operator thus falls into a particular case of the corresponding Lie-superalgebraic quasi-solvable model. In addition to the two models treated in this paper, there exist other Lie-superalgebraic quasi-solvable matrix models such as the $\fosp(2/1)$ ones~\cite{SK97}, the $\fq(2)$ ones realized by the first and the third representations in~\cite{DJ01}, and so on. Application of $\cN$-fold SUSY to them would further reveal the mathematical structure underlying between quasi-solvability and $\cN$-fold SUSY.

It should be noted that the framework of $\cN$-fold SUSY has various advantageous aspects superior to the Lie-superalgebraic approach. First, it enables one to construct not only a quasi-solvable operator but simultaneously a pair of such operators which share almost the same spectra. For the latter purpose, it is not necessary to impose the algebraic constraint (\ref{eq:Nf2}) since the intertwining (\ref{eq:Nf1}) solely guarantees it. Second, we recall the important fact that not all the quasi-solvable scalar systems so far discovered are available by means of Lie-algebraic generators; recent development have suggested that Lie-algebraic systems are rather exceptional, see e.g.,~\cite{Ta13}. Needless to say, the $\cN$-fold SUSY formulation does not rely on any such algebraic tools and works well also in constructing quasi-solvable systems which are not expressible by a representation of certain kind of algebra. Hence, just by choosing a suitable form of $\tbP_{\cN}^{-}$, we would be able to obtain new quasi-solvable matrix systems which are free from Lie superalgebra.

Finally, coming back to the models in this paper, we refer to some remaining issues on them to be followed in the future, in particular, on the $\fosp(2/2)$ model, since the structure of the $\fq(2)$ model has been already virtually uncovered. To appreciate the relation between the $\fosp(2/2)$ model and the corresponding $\cN$-fold SUSY one, it is necessary to obtain the full solution to the intertwining relation for $\cN\geq 3$ where the former gets entirely Lie superalgebraic and all the coefficients become polynomials in $z$. The role played by $\beta(z)$ in the case of $\alpha(z)\neq 0$ has not been clear in this context and it is obviously more desirable to solve the intertwining relation by including it. We would also like to know what kind of quantum mechanical matrix models are realized in the physical $q$-space; in this paper we have mostly worked in the gauged $z$-space where we can analyze the mathematical structure more transparently and systematically. Classification of the models in the $q$-space is another important remaining problem.

\begin{acknowledgments}
This work was supported in part by JSPS KAKENHI Grant Number 26400386.
\end{acknowledgments}



\bibliography{refsels}
\bibliographystyle{npb}



\end{document}